%% file: TPDS_vlsi_submission_v0.tex
\documentclass[10pt,journal,compsoc]{IEEEtran}

\usepackage[justification=centering]{caption}
\usepackage{algorithm,algpseudocode}
\usepackage{amsmath}
\usepackage{graphicx}
\usepackage{times}                               % LaTeX 2e
\usepackage{graphicx}
\usepackage{tabularx}
\usepackage{array}
\usepackage{amsmath}
\usepackage{amssymb}
\usepackage{units}
\usepackage{textcomp,xspace}
\usepackage{enumerate}
\usepackage{multirow}
\usepackage{multicol}
\usepackage{subfigure}
\usepackage{color}
\usepackage{varwidth}

\makeatletter
\def\BState{\State\hskip-\ALG@thistlm}
\makeatother

\algdef{SE}[DOWHILE]{Do}{doWhile}{\algorithmicdo}[1]{\algorithmicwhile\ #1}

\newcommand{\R}{\mathbb{R}}

\ifCLASSINFOpdf

\else

\fi

\begin{document}

\title{Efficient Realization of Givens Rotation through Algorithm-Architecture Co-design for Acceleration of QR Factorization}
%\title{Achieving Efficient QR Factorization by Algorithm-Architecture Co-Design of Householder Transformation}

\author{Farhad Merchant,
        Tarun Vatwani, Anupam Chattopadhyay, ~\IEEEmembership{Senior Member, IEEE,} Soumyendu Raha, \\S K Nandy, ~\IEEEmembership{Senior Member, IEEE,}, Ranjani Narayan, and Rainer Leupers
        % <-this % stops a space
\IEEEcompsocitemizethanks{\IEEEcompsocthanksitem Farhad Merchant and Rainer Leupers are with Institute for Communication Technologies and Embedded Systems, RWTH Aachen University, Germany
Email:\{farhad.merchant, leupers\}@ice.rwth-aachen.de
\IEEEcompsocthanksitem Tarun Vatwani and Anupam Chattopadhyay are with School of Computer Science and Engineering,
Nanyang Technological University, Singapore\protect\\
E-mail: \{tvatwani,anupam\}@ntu.edu.sg
\IEEEcompsocthanksitem Soumyendu Raha and S K nandy are with Indian Institute of Science, Bangalore \IEEEcompsocthanksitem Ranjani Narayan is with Morphing Machines Pvt. LTd. }% <-this % stops a space
\thanks{Manuscript received October 20, 2016;}}

%\author{Farhad Merchant,
%        Tarun Vatwani, Anupam Chattopadhyay, ~\IEEEmembership{Senior Member, IEEE,} Soumyendu Raha, \\S K Nandy, ~\IEEEmembership{Senior Member, IEEE,} and Ranjani Narayan
%        % <-this % stops a space
%\IEEEcompsocitemizethanks{\IEEEcompsocthanksitem Farhad Merchant, Tarun Vatwani, and Anupam Chattopadhyay are with School of Computer Science and Engineering,
%Nanyang Technological University, Singapore\protect\\
%% note need leading \protect in front of \\ to get a newline within \thanks as
%% \\ is fragile and will error, could use \hfil\break instead.
%E-mail: \{mamirali,tarun,anupam\}@ntu.edu.sg
%%\IEEEcompsocthanksitem Tarun Vatwani is with Indian Institute of Technology, Jodhpur 
%\IEEEcompsocthanksitem Soumyendu Raha and S K nandy are with Indian Institute of Science, Bangalore \IEEEcompsocthanksitem Ranjani Narayan is with Morphing Machines Pvt. LTd. }% <-this % stops a space

% \markboth{Journal of \LaTeX\ Class Files,~Vol.~14, No.~8, August~2015}%
% {Shell \MakeLowercase{\textit{et al.}}: Bare Demo of IEEEtran.cls for IEEE Journals}

\IEEEtitleabstractindextext{%
\begin{abstract}
\input{abstract}
\end{abstract}

\begin{IEEEkeywords}
Parallel computing, orthogonal transforms, dense linear algebra, multiprocessor system-on-chip, instruction level parallelism
\end{IEEEkeywords}}

\maketitle

\IEEEpeerreviewmaketitle

\section{Introduction}
\input{introduction}

\section{Background and Related Work}\label{sec:rw_back}
\input{related_work}

\section{Case Studies}\label{sec:casestudy}
\input{case_studies}

\section{Generalized Givens Rotation and Implementation}\label{sec:imp}
\input{implementation}

\section{Parallel Implementation of GGR in REDEFINE}\label{sec:parallelimp}
\input{parallel_imp}

\section{Conclusion}\label{sec:con}
\input{conclusion}

%%\section{Introduction}
%%\label{sec:intro}
%%\input{intro_new}
%%\section{Background and Related Work}
%%\label{sec:rw}
%%\input{related_work}
%%\section{Case Studies}\label{sec:mot}
%%\input{case_study}
%%\section{Modified Householder Transform}\label{sec:ht}
%%\input{house}
%%\section{Custom Realization of Householder Transform and Results}\label{sec:cus}
%%\input{custom}
%%%\section{Results}\label{sec:res}
%%
%%\section{Conclusion}\label{sec:con}
%%\input{conclusion}

\bibliographystyle{IEEEtran}
\bibliography{IEEEabrv,ref}

\begin{IEEEbiography}[{\includegraphics[width=1in,height=1.25in,clip,keepaspectratio]{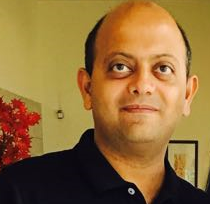}}]{Farhad Merchant} is a Postdoctoral Research Fellow at Institute for Communication Technologies and Embedded Systems, RWTH Aachen University, Germany. Previously he has worked as a Researcher at Research and Technology Center, Robert Bosch and before that he was Research Fellow at Hardware and Embedded Systems Lab, School of Computer Science and Engineering, Nanyang Technological University, Singapore. He received his PhD from Computer Aided Design Laboratory, Indian Institute of Science, Bangalore, India. His research interests are algorithm-architecture co-design, computer architecture, reconfigurable computing, development and tuning of high performance software packages
\end{IEEEbiography}

\begin{IEEEbiography}[{\includegraphics[width=1in,height=1.25in,clip,keepaspectratio]{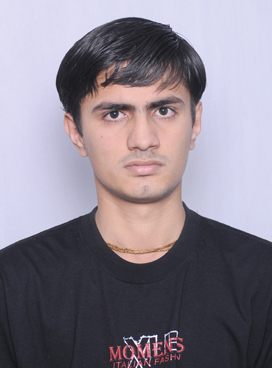}}]{Tarun Vatwani}
is a Research Associate at Hardware and Embedded Systems Lab, School of Computer Science and Engineering, Nanyang Technological University, Singapore. He is a B.Tech. graduate from Indian Institute of Technology, Jodhpur, India, His research interests are computer architecture, high performance computing, machine learning, performance tuning of different software packages.
\end{IEEEbiography}

\begin{IEEEbiography}[{\includegraphics[width=1in,height=1.25in,clip,keepaspectratio]{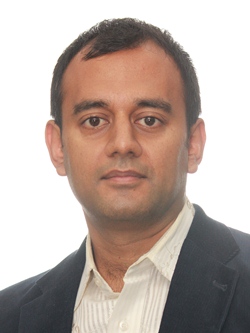}}]{Anupam Chattopadhyay}
 received his B.E. degree from Jadavpur University, India in 2000. He received his MSc. from ALaRI, Switzerland and PhD from RWTH Aachen in 2002 and 2008 respectively. From 2008 to 2009, he worked as a Member of Consulting Staff in CoWare R\&D, Noida, India. From 2010 to 2014, he led the MPSoC Architectures Research Group in RWTH Aachen, Germany as a Junior Professor. Since September, 2014, he is appointed as an assistant Professor in SCE, NTU.
\end{IEEEbiography}
\vspace{-10mm}
\begin{IEEEbiography}[{\includegraphics[width=1in,height=1.25in,clip,keepaspectratio]{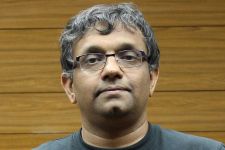}}]{Soumyendu Raha}
obtained his PhD in Scientific Computation from the University of Minnesota in 2000.
Currently he is a Professor of the Computational and Data Sciences Department at the Indian
Institute of Science in Bangalore, which he joined in 2003, after having worked for IBM for a couple of years.
His research interests are in computational mathematics of dynamical systems, both continuous and combinatorial,
and in co-development and application of computing systems for implementation of computational mathematics algorithms.
\end{IEEEbiography}

\begin{IEEEbiography}[{\includegraphics[width=1in,height=1.25in,clip,keepaspectratio]{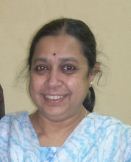}}]
{Ranjani Narayan} has over 15 years experience at
IISc and 9 years at Hewlett Packard. She has vast work
experience in a variety of fields – computer architecture,
operating systems, and special purpose systems. She
has also worked in the Technical University of Delft, The
Netherlands, and Massachusetts Institute of Technol-
ogy, Cambridge, USA. During her tenure at HP, she
worked on various areas in operating systems and
hardware monitoring and diagnostics systems. She has
numerous research publications.She is currently Chief
Technology Officer at Morphing Machines Pvt. Ltd,
Bangalore, India.
\end{IEEEbiography}

\begin{IEEEbiography}[{\includegraphics[width=1in,height=1.25in,clip,keepaspectratio]{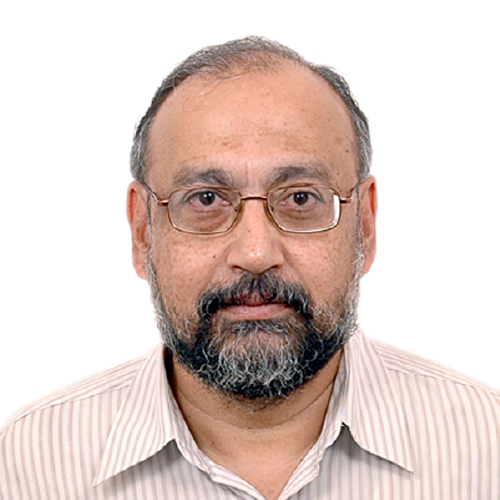}}]{S K Nandy}
is a Professor in the Department of Computational and Data Sciences of the Indian Institute of Science, Bangalore.
His research interests are in areas of High Performance Embedded Systems on a  Chip, VLSI architectures for Reconfigurable Systems on Chip,
and Architectures and Compiling Techniques for Heterogeneous Many Core Systems. Nandy received the B.Sc (Hons.) Physics degree from the Indian Institute of Technology, Kharagpur, India, in 1977. He obtained the BE (Hons.) degree
 in Electronics and Communication in 1980, MSc.(Engg.) degree in Computer Science and Engineering in 1986, and the Ph.D. degree in Computer Science
and Engineering in 1989 from the Indian Institute of Science, Bangalore. He has over 170 publications in International Journals, and Proceedings of International Conferences, and 5 patents.
\end{IEEEbiography}

\begin{IEEEbiography}[{\includegraphics[width=1in,height=1.25in,clip,keepaspectratio]{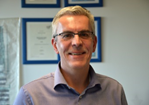}}]{Rainer Leupers}
received the M.Sc. (Dipl.-Inform.) and Ph.D. (Dr. rer. nat.) degrees in Computer Science with honors from TU Dortmund in 1992 and 1997. From 1997-2001 he was the chief engineer at the Embedded Systems chair at TU Dortmund. In 2002, he joined RWTH Aachen University as a professor for Software for Systems on Silicon. His research comprises software development tools, processor architectures, and system-level electronic design automation, with focus on application-specific multicore systems. He published numerous books and technical papers and served in committees of the leading international EDA conferences. He received various scientific awards, including Best Paper Awards at DAC and twice at DATE, as well as several industrial awards. Dr. Leupers is also engaged as an entrepreneur and in turning novel technologies into innovations. He holds several patents on system-on-chip design technologies and has been a co-founder of LISATek (now with Synopsys), Silexica, and Secure Elements. He has served as consultant for various companies, as an expert for the European Commission, and in the management boards of large-scale projects like HiPEAC and UMIC. He is the coordinator of EU projects TETRACOM and TETRAMAX on academia/industry technology transfer.
\end{IEEEbiography}

\end{document}

%% file: abstract.tex
We present efficient realization of Generalized Givens Rotation (GGR) based QR factorization that achieves 3-100x better performance in terms of Gflops/watt over state-of-the-art realizations on multicore, and General Purpose Graphics Processing Units (GPGPUs). GGR is an improvement over classical Givens Rotation (GR) operation that can annihilate multiple elements of rows and columns of an input matrix simultaneously. GGR takes 33\% lesser multiplications compared to GR. For custom implementation of GGR, we identify macro operations in GGR and realize them on a Reconfigurable Data-path (RDP) tightly coupled to pipeline of a Processing Element (PE). In PE, GGR attains speed-up of 1.1x over Modified Householder Transform (MHT) presented in the literature. For parallel realization of GGR, we use REDEFINE, a scalable massively parallel Coarse-grained Reconfigurable Architecture, and show that the speed-up attained is commensurate with the hardware resources in REDEFINE. GGR also outperforms General Matrix Multiplication (gemm) by 10\% in-terms of Gflops/watt which is counter-intuitive. 

%% file: introduction.tex
\begin{figure}[!ht]
	\centering
	\includegraphics[scale = 0.27]{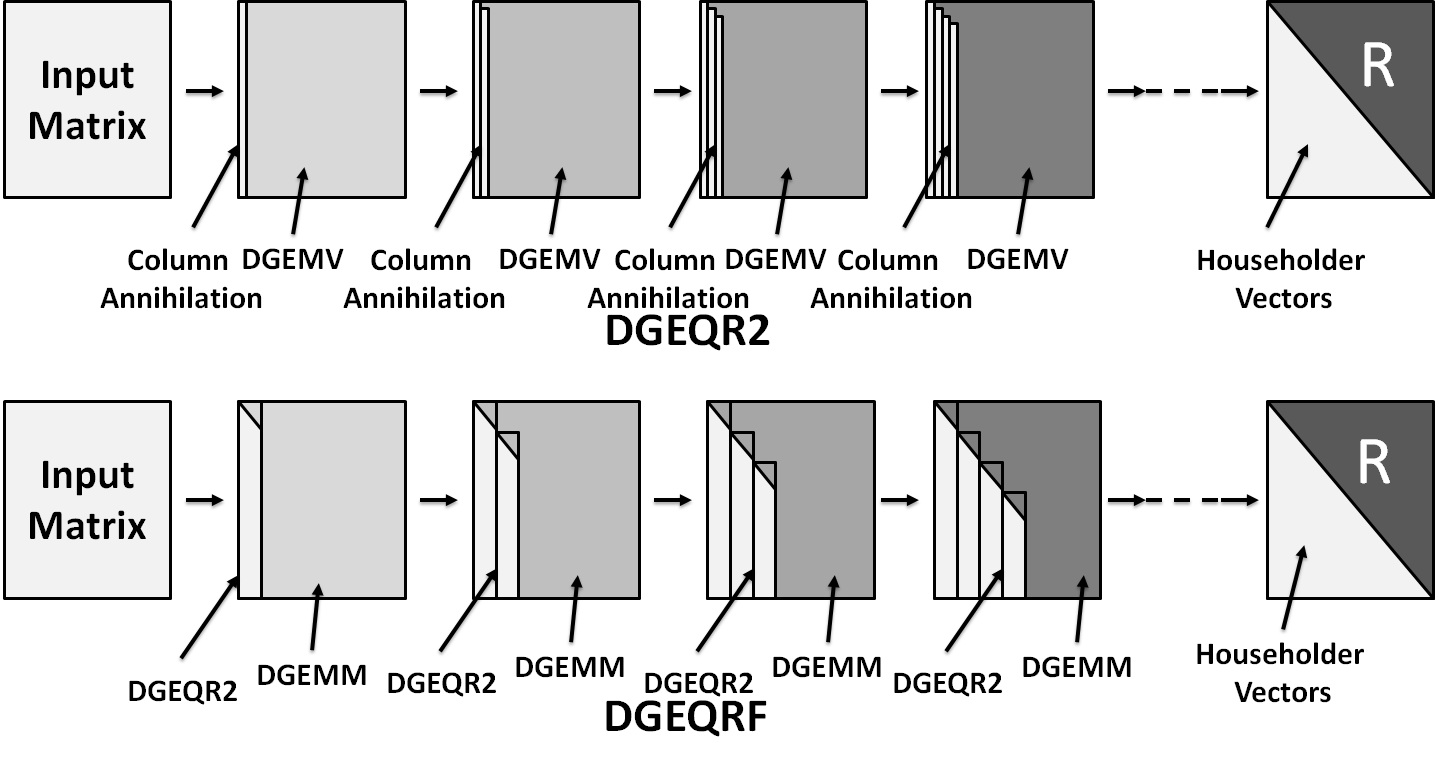}
	\caption{dgeqr2 and dgeqrf Operations}
	\label{fig:fig1}
\end{figure}

QR factorization/decomposition (QRF/QRD) is a prevalent operation encountered in several engineering and scientific operations ranging from Kalman Filter (KF) to computational finance \cite{kal_new1}\cite{far_kal1}. 
QR factorization of a non-singular matrix $A_{m\times n}$ of size $m\times n$ is given by 
\begin{align}\label{eqn:qrd1}
	A = QR
\end{align}
where $Q_{m\times m}$ is an orthogonal and $R_{m\times n}$ is upper triangular matrix. There are mainly three methods to compute QR factorization, 1) Householder Transform (HT), 2) Givens Rotation (GR), and 3) Modified Gram-Schmidt (MGS). MGS is used in the embedded systems where numerical accuracy of the final solution is not critical, while HT is employed in High Performance Computing (HPC) applications since it is numerically stable operation. GR is applied in the application domains pertaining to embedded systems where numerical stability of the end solution is critical \cite{Min1}. We sense here an opportunity in GR for applicability in domains beyond embedded systems. Specifically, we foresee opportunity in GR in generalization for annihilation of multiple elements of an input matrix simultaneously where annihilation regime spans over columns and rows. It is intended to expose higher parallelism in classical GR through generalization and also reduction in total number computations in GR. Such a generalization is possible by combining several Givens sequences and performing common computations required for updating trailing matrix beforehand. Surprisingly, such an approach is nowhere presented in the literature and that has reduced relevance of GR in several application domains. It has been emphasized in the literature that GR is more suitable for orthogonal decomposition of sparse matrices while for orthogonal decomposition of dense matrices, HT is more suitable over GR \cite{spgr1}\cite{alan1}.  For implementations on multicore and General Purpose Graphics Processing Units (GPGPUs), a library based approach is followed for Dense Linear Algebra (DLA) computations where highly tuned packages based on specifications given in Basic Linear Algebra Subprograms (BLAS) and Linear Algebra Package (LAPACK) are developed \cite{lapack1}. Several realizations of QR factorization are discussed in section \ref{sec:rw}. Typically, block QR factorization routine (xgeqrf - where x indicates double/single precision) that is part of LAPACK is realized as a series of BLAS calls. dgeqr2 and dgeqrf operations are shown in figure \ref{fig:fig1}. 
In dgeqr2, Double Precision Matrix-vector (dgemv) operation is dominant while in dgeqrf, Double Precision Matrix Multiplication (dgemm) is dominant as depicted in the figure \ref{fig:fig1}. dgemv can attain up to 10\% of the theoretical peak in multicore platforms and 15-20\% in GPGPUs while dgemm can attain up to 40-45\% in multicore platforms and 55-60\% in GPGPUs at $\approx 65W$ and $\approx 260W$ respectively \cite{tpds1}. Considering low performance of multicores and GPGPUs for critical DLA computations, it could be a prudent approach to move away from traditional BLAS/LAPACK based strategy in software and accelerate these computations on a customizable platform that can attain order of magnitude higher performance than state-of-the-art realizations of these software packages. A special care has to be taken in designing of an accelerator that is capable of attaining desired performance while maintaining generality of the accelerator in also supporting other operations in the domain of DLA computations. Coarse-grained Reconfigurable Architectures (CGRAs) are a good candidate for the domain of DLA computations since they are capable of attaining performance of Application Specific Integrated Circuits (ASICs) while flexibility of Field Programmable Gate Arrays (FPGAs) \cite{derek1}\cite{cgr2}\cite{cgr3}. Recently, there have been several proposals in the literature in developing BLAS and LAPACK on custamizable CGRA platforms through algorithm-architecture co-design where macro operations in the operations pertaining to DLA computations are identified and realized on a Reconfigurable Data-path (RDP) that is tightly coupled to the processor pipeline \cite{hyper1}\cite{egra1}. In this paper, we focus on acceleration of GR based QR factorization, where classical GR is generalized to achieve Generalized Givens Rotation (GGR) where GGR has 33\% lesser multiplications compared to GR. Several macro operations in GGR are identified and realized on RDP to achieve superior performance compared to Modified Householder Transform (MHT) presented in \cite{tpds1}. Major contributions in this paper to achieve efficient realization of GR based QR factorization are as follows:
\begin{itemize}
	\item We improvise over Column-wise Givens Rotation (CGR) presented in \cite{cgr1} and present GGR. While CGR is capable of simultaneous annihilation of multiple elements of a column in the input matrix, GGR can annihilate multiple elements of rows and columns simultaneously
	\item Several macro operations in GGR are identified and implemented on an RDP that is tightly coupled to pipeline of a Processing Element (PE) resulting in 81\% of the theoretical peak in PE. This implementation outperforms Modified Householder Transform (MHT) based QR factorization (dgeqr2ht) implementation presented in \cite{tpds1} by 10\% in PE. GGR based QR factorization also outperforms dgemm in PE by 10\% which is counter-intuitive
	\item Arguably, moving away from BLAS for realization of GGR based QR factorization attains 10\% higher performance than the classical way of implementation where Level-3 BLAS is used as a dominant operation in the state-of-the-art software packages for multicore and GPGPUs. This claim is validated by several case studies on multicore and GPGPUs where it is also shown that moving away from BLAS and LAPACK in these platforms does not yield performance improvement
	\item For parallel realization in REDEFINE, we attach PE in REDEFINE framework where 10\% higher performance is attained over dgeqr2ht implementation presented in \cite{tpds1}. We show that sequential realization in PE and parallel realization of GGR based QR factorization in REDEFINE are scalable. Furthermore, it is shown that the speed-up in parallel realization in REDEFINE over sequential realization in PE is commensurate with the hardware resources employed in REDEFINE and the speed-up asymptotically approaches theoretical peak of REDEFINE CGRA
\end{itemize}

For our implementations in PE and REDEFINE, we have used double precision Floating Point Unit (FPU) presented in \cite{fpu2} with recommendations presented in \cite{fpu3}. Organization of the papers is as follows: In section \ref{sec:rw_back}, we discuss about CGR, REDEFINE and some of the FPGA, multicore, and GPGPU based realizations of QR factorization. Case studies on dgemm, dgeqr2, dgeqrf, dgeqr2ht, and dgeqrfht are presented in section \ref{sec:casestudy}. GGR and implementation of GGR in multicore, GPGPU, and PE is discussed in \ref{sec:imp}. Parallel realization of GGR in REDEFINE CGRA is discussed in \ref{sec:parallelimp}. We conclude our work in section \ref{sec:con}.

\noindent {\bf Abbreviations/Nomenclature:}

{\scriptsize

\begin{center}
\begin{tabular}{ | m{9em} | m{5.5cm}| } 
\hline
Abbreviation/Name & Expansion/Meaning  \\ 
\hline \hline
AVX  & Advanced Vector Extension  \\ 
\hline
BLAS & Basic Linear Algebra Subprograms  \\ 
\hline
CE & Compute Element  \\ 
\hline
CFU & Custom Function Unit  \\ 
\hline
CGRA & Coarse-grained Reconfigurable Architecture  \\ 
\hline
CPI & Cycles-per Instruction  \\ 
\hline
CUDA & Compute Unified Device Architecture   \\ 
\hline
EREW-PRAM & Exclusive-read Exclusive-write Parallel Random Access Machine  \\ 
\hline
DAG & Directed Acyclic Graph  \\ 
\hline
FMA & Fused Multiply-Add  \\ 
\hline
FPGA & Field Programmable Gate Array  \\ 
\hline
FPS & Floating Point Sequencer  \\ 
\hline
FPU & Floating Point Unit \\ 
\hline
GPGPU & General Purpose Graphics Processing Unit  \\ 
\hline
GR & Givens Rotation  \\ 
\hline
GGR & Generalized Givens Rotation \\
\hline
HT & Householder Transform  \\ 
\hline
ICC & Intel C Compiler  \\ 
\hline
IFORT & Intel Fortran Compiler  \\ 
\hline
ILP & Instruction Level Parallelism  \\ 
\hline
KF & Kalman Filter  \\ 
\hline
LAPACK & Linear Algebra Package  \\ 
\hline
MAGMA & Matrix Algebra on GPU and Multicore Architectures  \\ 
\hline
MGS & Modified Gram-Schmidt  \\ 
\hline
MHT & Modified Householder Transform  \\ 
\hline
NoC & Network-on-Chip \\
\hline
PE & Processing Element  \\ 
\hline
PLASMA & Parallel Linear Algebra Software for Multicore Architectures  \\ 
\hline
QUARK & Queuing and Runtime for Kernels  \\ 
\hline
RDP & Reconfigurable Data-path  \\ 
\hline
XGEMV/xgemv & Single/Double Precision General Matrix-vector Multiplication  \\ 
\hline
XGEMM/xgemm & Single/Double Precision General Matrix Multiplication  \\ 
\hline
XGEQR2/dgeqr2 & Single/Double Precision QR Factorization based on Householder Transform (with XGEMV)  \\  \hline
XGEQRF/xgeqrf & Blocked Single/Double Precision QR Factorization based on Householder Transform (with XGEMM)  \\ \hline
XGEQR2HT/xgeqr2ht & Single/Double Precision QR Factorization based on Modified Householder Transform  \\ \hline
XGEQR2GGR/ xgeqr2ggr & Single/Double Precision QR Factorization based on Generalized Givens Rotation  \\ \hline
XGEQRFHT/xgeqrfht & Blocked Single/Double Precision QR Factorization based on Modified Householder Transform (with XGEMM)  \\ \hline
XGEQRFGGR/ xgeqrfggr & Blocked Single/Double Precision QR Factorization based on Generalized Givens Rotation (with XGEMM)  \\ \hline
PACKAGE\_ROUTINE/ package\_routine & Naming convention followed for different routines pertaining to different packages. E.g., BLAS\_DGEMM/ blas\_dgemm is a Double Precision General Matrix Multiplication Routine in BLAS  \\ 
\hline
\end{tabular}
\end{center}
}

%% file: related_work.tex
CGR presented in \cite{cgr1} is discussed in section \ref{sec:cgr} and REDEFINE CGRA is discussed in section \ref{sec:redefinecgra}. A detailed review of yesteryear realizations of QR factorization is presented in section \ref{sec:rw}. 
\subsection{Givens Rotation based QR Factorization}\label{sec:cgr}
For a $4\times 4$ matrix $X=x_{ij},x_{ij}\in \R^{4\times 4}$, applying $3$ Givens sequences simultaneously yields to the matrix $GX$ shown in equation \ref{eqn:cgr_4}.
\begin{figure*}[!ht]
	\centering
	\includegraphics[scale = 0.35]{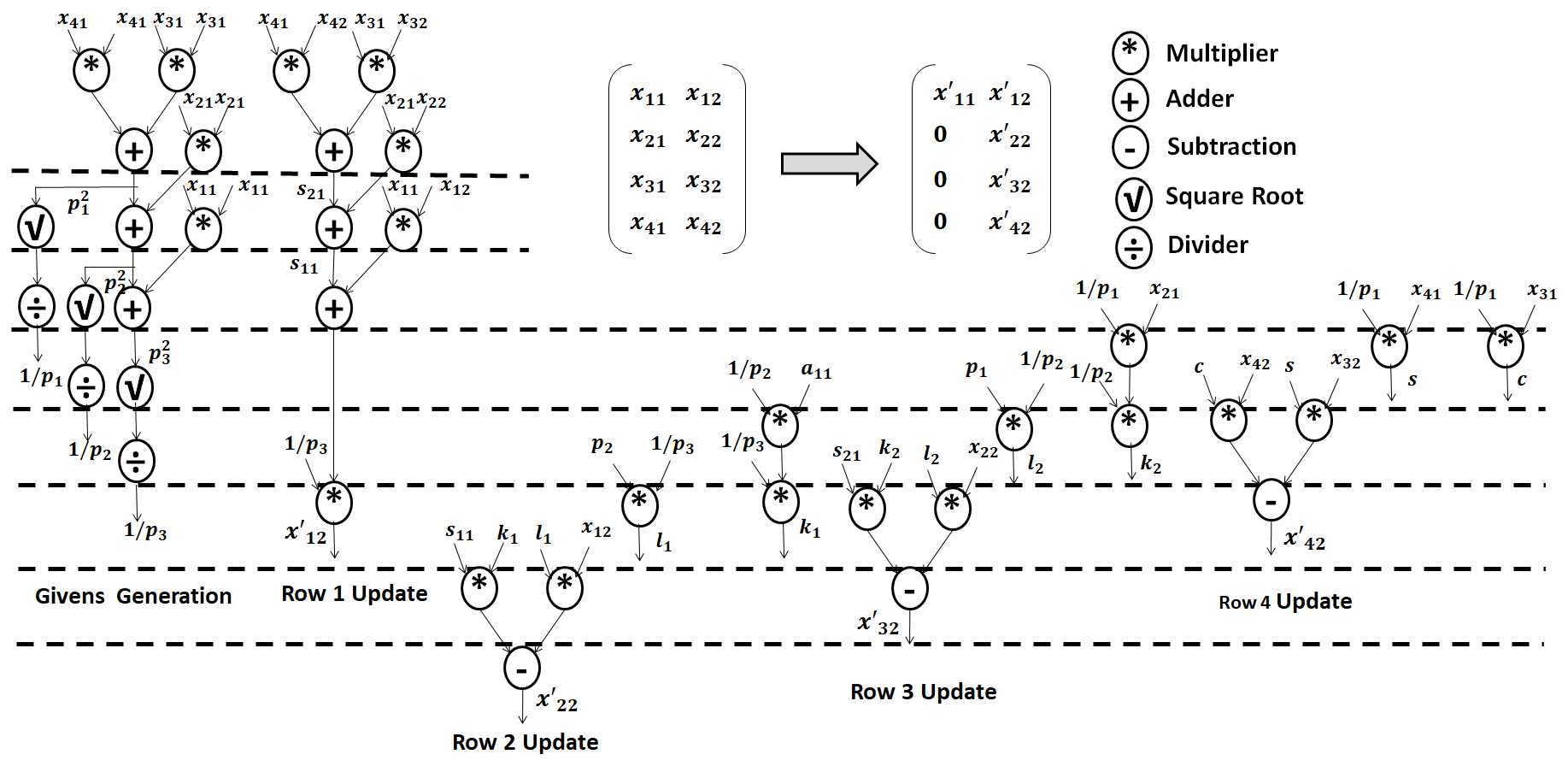}
	\caption{One Iteration of Column-wise Givens Rotation}
	\label{fig:fig2}
\end{figure*}
\begin{align} 
 & GX  = \nonumber \\ & \begin{bmatrix} 
p_3 & \frac{x_{11}x_{12} + s_{11}}{p_3} & \frac{x_{11}x_{13}+s_{12}}{p_3} & \frac{x_{11}x_{14} + s_{13}}{p_3} \\ 0 & \frac{x_{11}s_{11}}{p_3p_2} - \frac{x_{12}p_2}{p_3} & \frac{x_{11}s_{12}}{p_3p_2} - \frac{x_{13}p_2}{p_3} & \frac{x_{11}s_{13}}{p_3p_2} - \frac{x_{14}p_2}{p_3} \\
0  & \frac{x_{21}s_{21}}{p_2p_1} - \frac{x_{22}p_1}{p_2} & \frac{x_{21}s_{22}}{p_2p_1} - \frac{x_{23}p_1}{p_2} & \frac{x_{21}s_{23}}{p_2p_1} - \frac{x_{24}p_1}{p_2}  \\ 0 & \frac{x_{31}x_{42}}{p_1} - \frac{x_{41}x_{32}}{p_1} & \frac{x_{31}x_{43}}{p_1} - \frac{x_{41}x_{33}}{p_1} & \frac{x_{31}x_{44}}{p_1} - \frac{x_{41}x_{34}}{p_1}\end{bmatrix}  \nonumber \\ 
\label{eqn:cgr_4}
& = \begin{bmatrix}   p_3  & \frac{x_{11}x_{12} + s_{11}}{p_3} & \frac{x_{11}x_{13}+s_{12}}{p_3} & \frac{x_{11}x_{14} + s_{13}}{p_3} \\ 0 & k_{1}s_{11} - x_{12}l_{1} & k_ 1s_{12} - x_{13}l_1 & k_1s_{13} - x_{14}l_1  \\ 0 & k_{2}s_{21} - x_{22}l_{2} & k_ 2s_{22} - x_{23}l_2 & k_2s_{23} - x_{24}l_2  \\
0  & cx_{42} - x_{32}s  & cx_{43} - x_{33}s  & cx_{44} - x_{34}s\end{bmatrix}
\end{align} 
Corresponding Directed Acyclic Graphs (DAGs) for CGR for annihilation of $x_{41}$, $x_{31}$, and $x_{21}$ and update of the second column of the matrix $X$ are shown in figure \ref{fig:fig1}. For an input matrix of size $n\times n$ classical GR takes $\frac{n(n-1)}{2}$ sequences while CGR takes $n-1$ sequences. Furthermore, if the number of multiplications in GR is $GR_M$ and number of multiplications in CGR is $CGR_M$ then
\begin{align}
\label{eqn:cgr_cc}
& CGR_M = \frac{2n^3+3n^2 - 5n}{2} \\
\label{eqn:gr_cc}
& GR_M =  \frac{4n^3-4n}{3}
\end{align} 
Taking ratio of equation \ref{eqn:cgr_cc} and \ref{eqn:gr_cc}
\begin{align}
\label{eqn:alpha}
\alpha = \frac{CGR_M}{GR_M} = \frac{3(2n+5)}{8(n+1)}
\end{align}
From equation \ref{eqn:alpha}, as $n\rightarrow \infty$, $\alpha \rightarrow \frac{3}{4}$. As we increase the size of the matrix, the number of multiplications in CGR asymptotically approaches to $\frac{3}{4}$ times the number of multiplications in GR. Implementation details of CGR on systolic array and in REDEFINE can be found in \cite{cgr1}.
\begin{figure}[!ht]
	\centering
	\includegraphics[scale = 0.15]{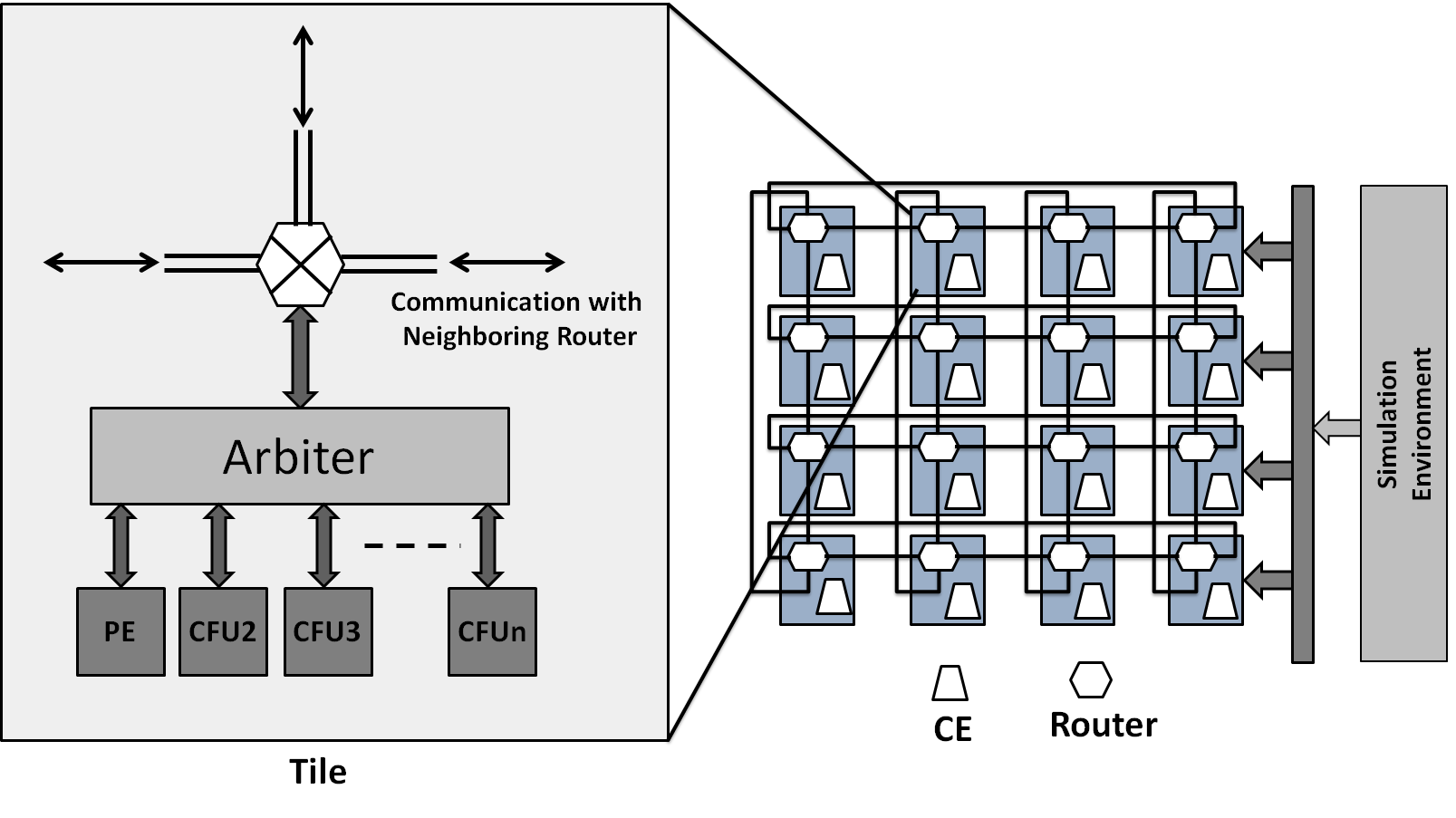}
	\caption{REDEFINE Framework}
	\label{fig:cfuredefine}
\end{figure}
\subsection{REDEFINE CGRA}\label{sec:redefinecgra}
REDEFINE CGRA is a customizable massively parallel Multiprocessor System on Chip (MPSoC) where several Tiles are connected through Network-on-Chip (NoC) \cite{Alle1}. Each Tile consists of a Compute Element (CE) and a Router. CEs in REDEFINE can be enhanced to support several applications domains like signal processing and Dense Linear Algebra (DLA) computations by attaching domain specific Custom Function Units (CFUs) to the CEs \cite{Merc1}. REDEFINE framework is shown in figure \ref{fig:cfuredefine}.
\begin{figure}%[!h]
\centering
\subfigure[Performance Comparison of PE with Other Platforms for dgemm and dgeqr2ht\label{fig:ht_graph_4}]{\includegraphics[scale = 0.15]{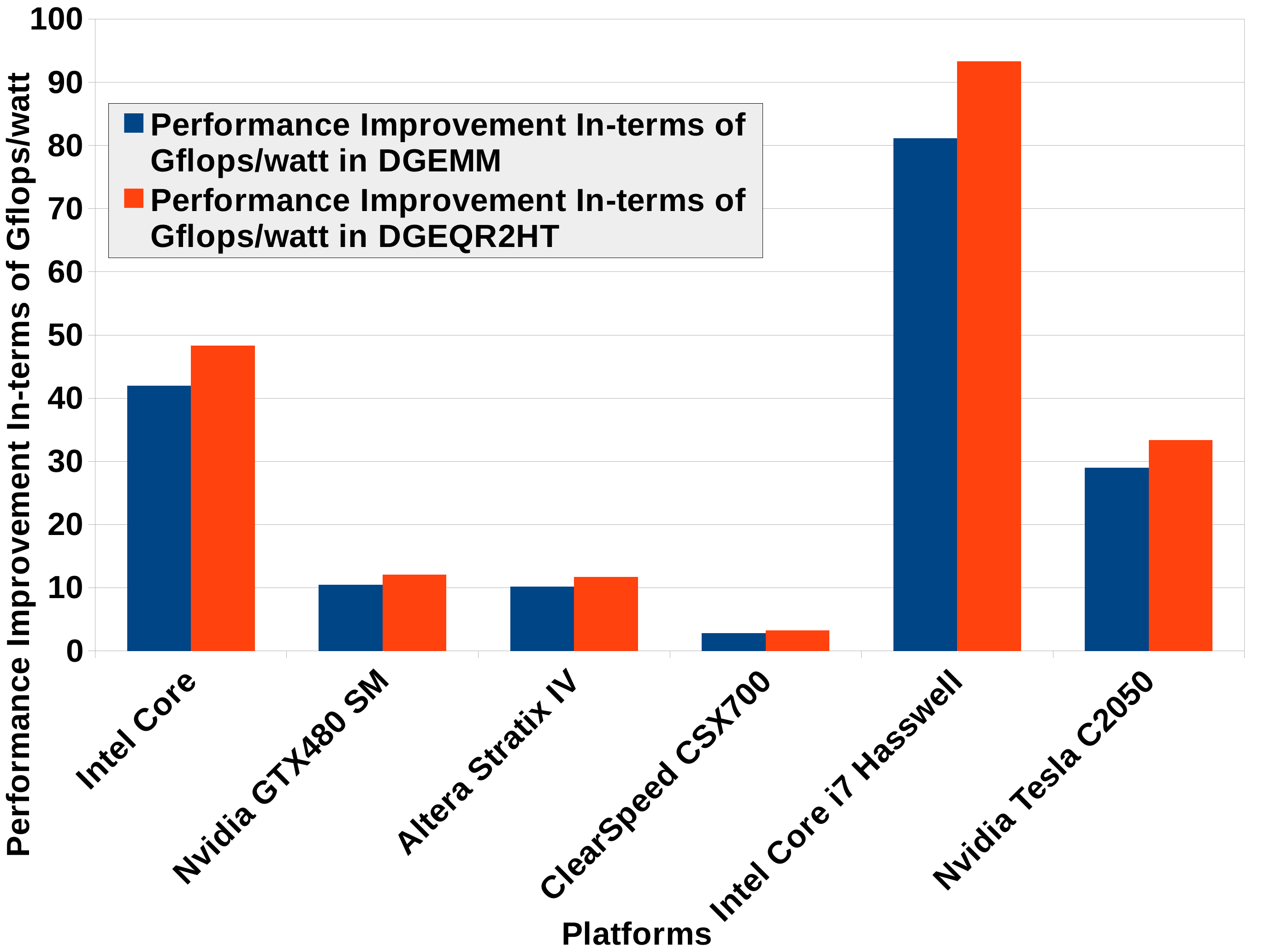}}
\subfigure[Performance of PE In-terms of Theoretical Peak Performance in PE for dgemm \label{fig:mm_explo5}]{\includegraphics[scale = 0.15]{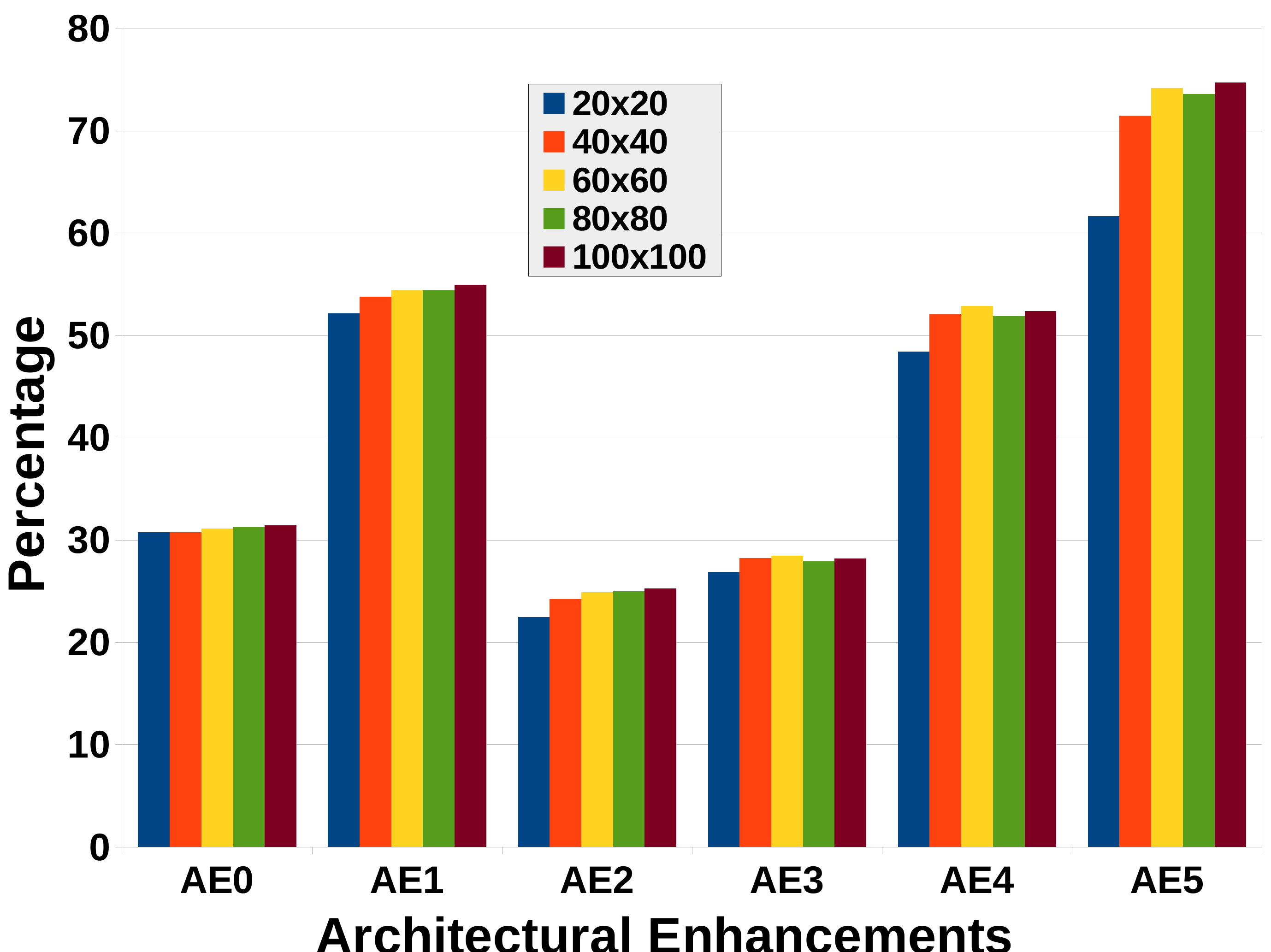}}
\caption{Performance and Performance Comparison of PE where dgeqr2ht is Modified Householder Transform based QR Factorization}
\label{fig:dgemm_perf}
\end{figure}
A Reconfigurbale Data-path is tightly coupled to a Processing Element (PE) that is CFU for REDEFINE as shown in the figure \ref{fig:cfuredefine}. Performance of PE in dgemm and MHT based QR factorization (dgeqr2ht) is shown in figure \ref{fig:ht_graph_4} \cite{tpds1}. Performance of PE over several Architectural Enhancements (AEs) is shown in figure \ref{fig:mm_explo5}. It can be observed in the figure \ref{fig:ht_graph_4} that PE attains 3-100x better performance in dgemm and dgeqr2ht while PE with tightly coupled RDP is capable of attaining 74\% of the theoretical peak performance of PE in dgemm as shown in figure \ref{fig:mm_explo5}. Performance in dgemm an dgeqr2ht is attained through algorithm-architecture co-design where macro operations in dgemm and dgeqr2ht are identified and realized on RDP. We apply similar technique in this exposition for GR where we first present GGR and identify macro operations in GGR that are realized on RDP. GGR implementation (dgeqr2ggr) outperforms dgeqr2ht and dgemm. Further details of dgemm and dgeqr2ht realizations can be found in \cite{Merc1}, \cite{exp1}, \cite{Merc2}, and \cite{tpds1}. 

\subsection{Related Work}\label{sec:rw}
Due to wide range of applications in the embedded systems domain, GR has been studied extensively in the literature specifically for implementation purpose since it was first proposed in \cite{gr_original}. An alternate ordering of Givens sequences was presented in \cite{Mod1}. According to the scheme presented in \cite{Mod1}, the alternate ordering is amenable to parallel implementation of GR while it does not focus on fusing several Givens sequences to annihilate multiple elements. For an input matrix of $n\times n$, the alternate ordering presented in \cite{Mod1} can annihilate maximum $\frac{n}{2}$ elements in parallel by executing disjoint Givens sequences simultaneously. Pipeline Given sequences for computing QR decomposition is presented in \cite{pipe} where its is proposed to execute Givens sequences in pipeline fashion to update the pair of rows partially updated by the previous Givens sequences. Analysis of this strategy is presented for Exclusive-read Exclusive-write Parallel Random Access Machine (EREW-PRAM) that shows that the pipeline strategy is twice as fast compared to the classical GR. Greedy Givens algorithm is presented in \cite{greedy} that executes compound disjoing Givens sequences in parallel assuming unlimited parallelism case. A high speed tournament GR and VLSI implementation of tournament GR is presented in \cite{Min1} where a significant improvement is reported in ASIC over triangular systolic array. The scheduling scheme presented in \cite{Min1} is similar to the one presented in \cite{Mod1} where disjoint Givens sequences are applied to compute QR decomposition. FPGA implementation of GR is presented in \cite{Aslan1} while ASIC implementation of square-root free GR is presented in \cite{Lei1}. A novel technique to avoid underflow/overflow in computation of QR decomposition using classical GR is presented in \cite{Jim1} that results in numerical stable realization of GR in LAPACK. A two-dimensional systolic array implementation of GR is presented in \cite{leeser1} where classical GR is implemented on two-dimensional systolic array with diagonal elements of the array performs complex operations like square root and division while the rest of the array performs matrix update. Restructuring of tridiagonal and bidiagonal algorithms for QR decomposition is presented in \cite{gr_fused1}. The restructuring strategy presented in \cite{gr_fused1} has several advantages like it is capable of exploiting vector instructions in the modern architectures, reduces memory operations, and the matrix updates are in the form of Level-3 BLAS thus capable of exploiting cache architecture through reuse of data compared to classical GR where it is Level-2 BLAS. Although, the scheme presented in \cite{gr_fused1} re-arranges memory bound operations like Level-2 BLAS in classical GR to compute bound operations like Level-3 BLAS, the scheme does not reduce computations unlike GGR. In general, works related to GR in the literature focus on different scheduling schemes for parallelism and exploitation of architectural features for the targeted platform. In case of GGR, total work is reduced compared to the classical GR while also being architecture platform friendly. For implementation of 
GR, there has been no work in the literature where macro operations in the routine are identified and realized carefully for performance. 

%% file: case_studies.tex
\begin{figure}%[!h]
\centering
\subfigure[PLASMA Software Stack\label{fig:plasma_soft}]{\includegraphics[scale = 0.12]{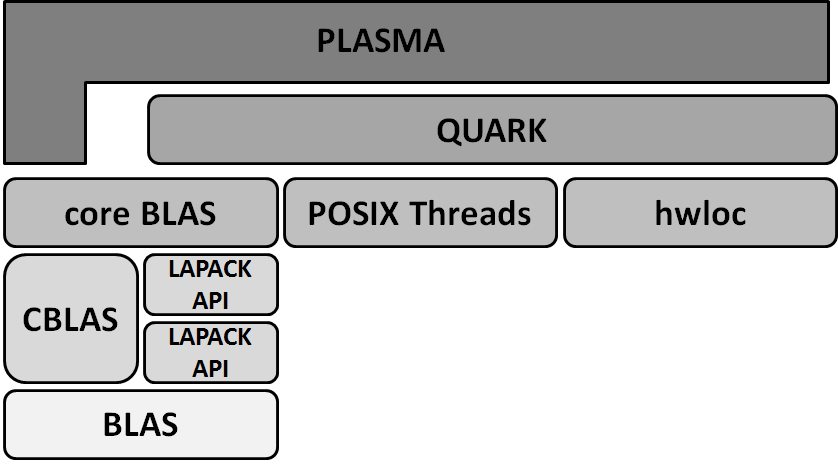}}
\subfigure[MAGMA Software Stack \label{fig:magma_soft}]{\includegraphics[scale = 0.12]{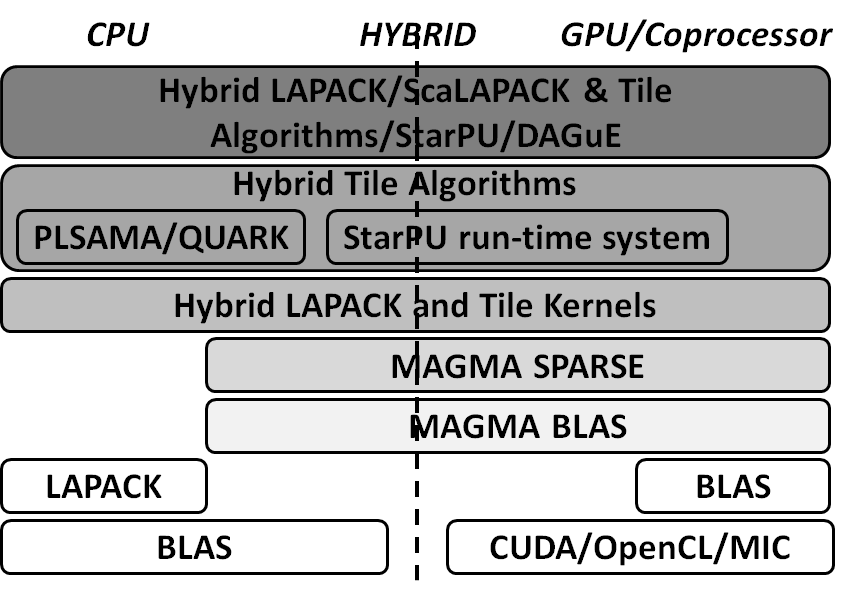}}
\caption{PLASMA and MAGMA Software Stacks}
\label{fig:plasma_magma}
\end{figure}
\begin{figure}%[!h]
\centering
\subfigure[Performance In-terms of Theoretical Peak Performance of Underlying Platform in dgemm, dgeqr2, and dgeqrf in LAPACK, PLASMA, and MAGMA\label{fig:cpi_gmm_7}]{\includegraphics[scale = 0.15]{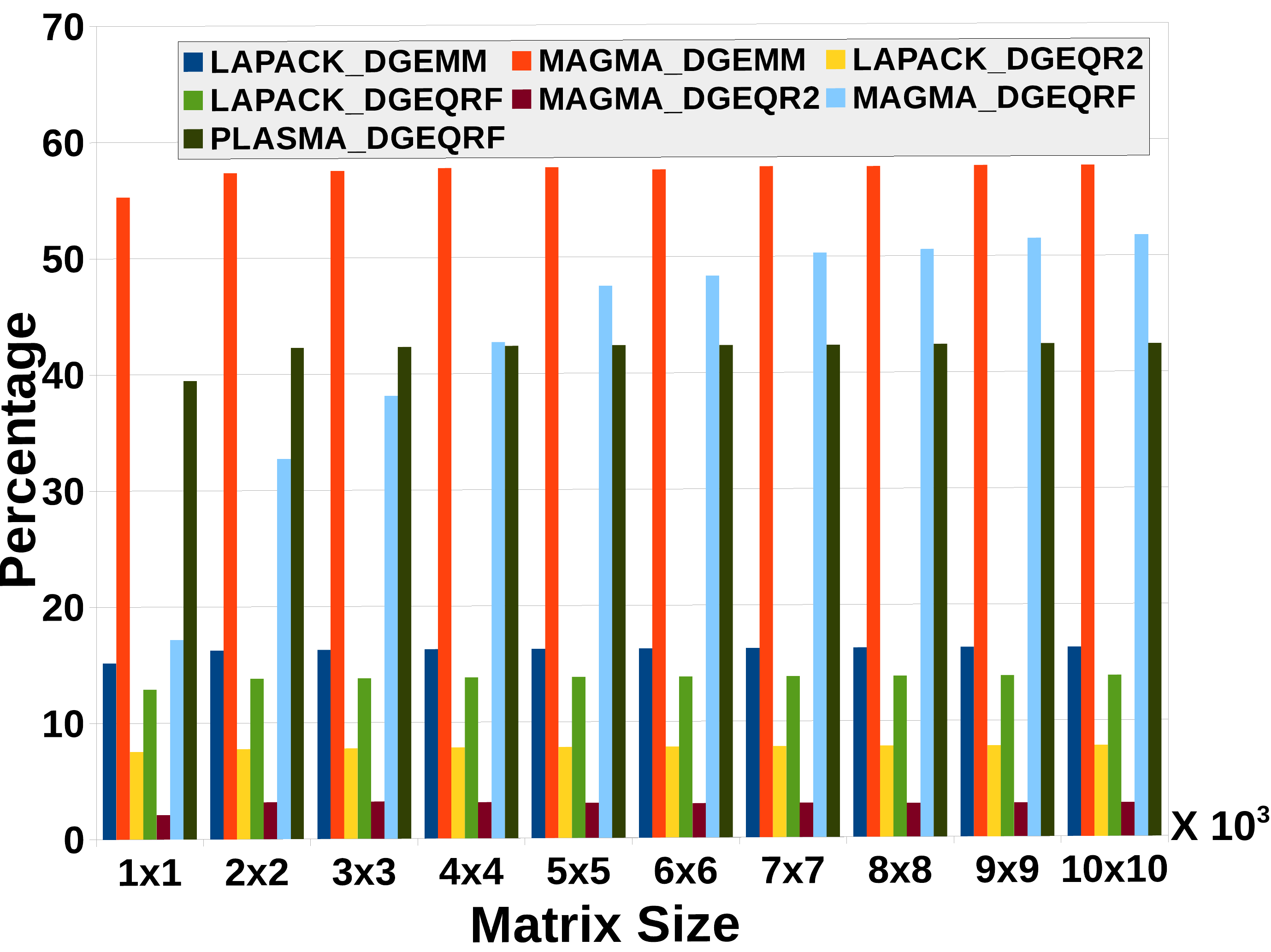}}
\subfigure[Performance In-terms of Gflops/watt in dgemm, dgeqr2, and dgeqrf in LAPACK, PLASMA, and MAGMA \label{fig:cpi_gmm_8}]{\includegraphics[scale = 0.15]{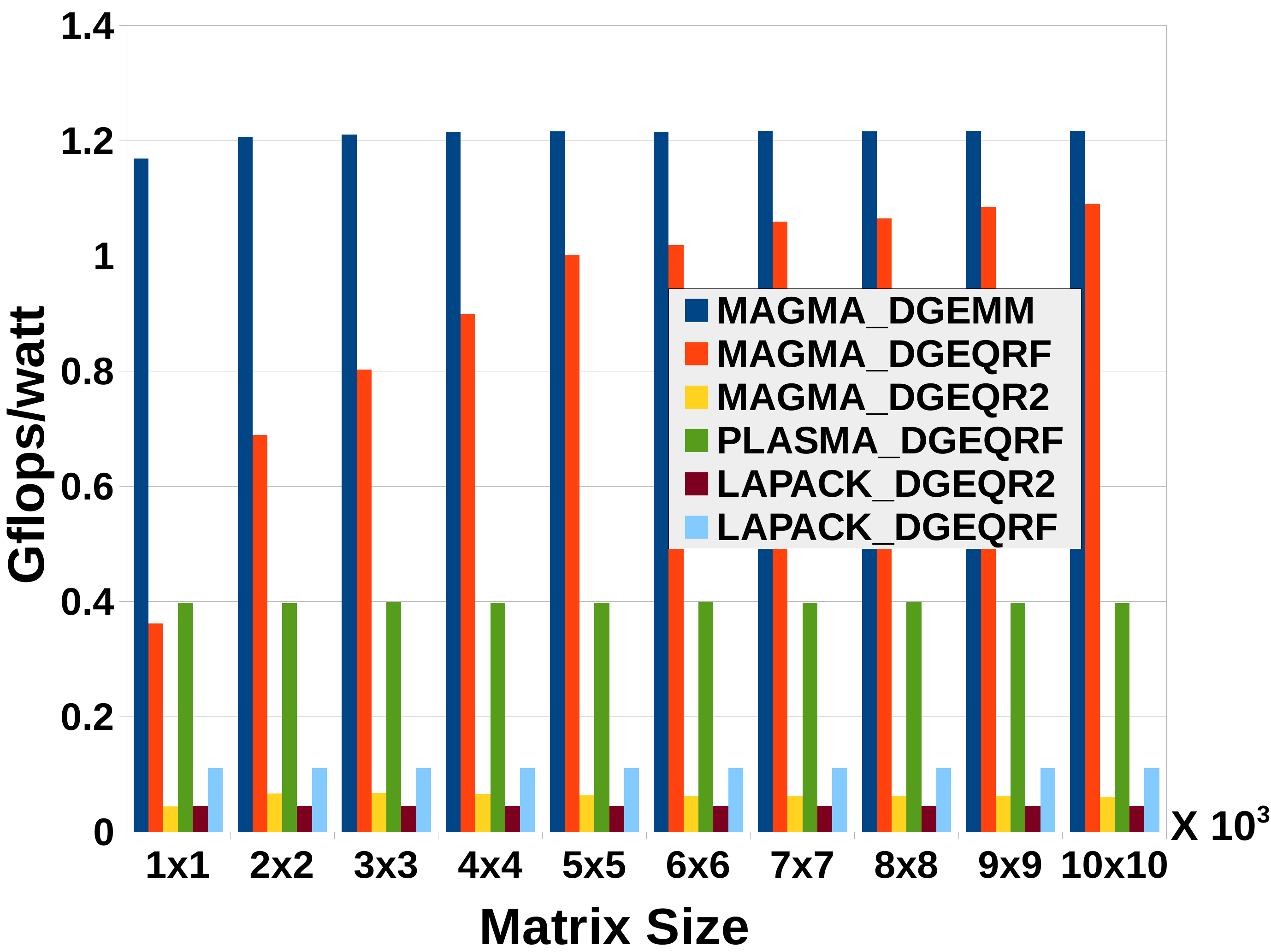}}
\caption{Performance in dgemm, dgeqr2, and dgeqrf in LAPACK, PLASMA, and MAGMA}
\label{fig:cpi_gmm}
\end{figure}

For our experiments on multicore and GPGPUs we use hightly tuned software packages PLASMA and MAGMA. Software stacks of these packages are shown in figure \ref{fig:plasma_magma}. 
Performance of PLASMA and MAGMA depends on efficiency of underlying BLAS realization as well as realization of scheduling schemes for multicore and GPGPUs. For implementation of GGR in PLASMA and MAGMA for multicore and GPGPUs, we add routines to BLAS and LAPACK. Performance in-terms of theoretical peak performance in dgemm, dgeqr2, and dgeqrf is depicted in figure \ref{fig:cpi_gmm_7} and performance in-terms of Gflops/watt for these routines is depicted in figure \ref{fig:cpi_gmm_8}. 

\subsection{dgemm}
dgemm is a Level-3 BLAS routine that has three loops and time complexity of $O(n^3)$ for a matrix of size $n\times n$. Performance of dgemm in LAPACK in-terms of theoretical peak of underlying platform is shown in figure \ref{fig:cpi_gmm_7} and in-terms of Gflops/watt in MAGMA is shown in figure \ref{fig:cpi_gmm_8}. It can be observed in the figure \ref{fig:cpi_gmm_7} that the performance attained by dgemm in Intel Core i7 and Nvidia Tesla C2050 is hardly 25\% and 57\% respectively. In-terms of Gflops/watt it is 1.22 in Nvidia Tesla C2050. Due to trivial nature of dgemm algorithm, we do not reproduce dgemm algorithm here while standard dgemm algorithm can be found in \cite{Golub1}. 
\subsection{dgeqr2}
\begin{algorithm}
\caption{dgeqr2 (Pseudo code)}
\label{algo:dgeqr21}
\begin{algorithmic}
\State{Allocate memory for input/output matrices and vectors}
\For{$i=1$ to $n$}
\State{Compute Householder vector $v$}
\State{Compute $P$ where $P = I - 2vv^T$}
\State{Update trailing matrix  using dgemv}
\EndFor
\end{algorithmic}
\end{algorithm}
Pseudo code of dgeqr2 is described in algorithm \ref{algo:dgeqr21}. It can be observed in the pseudo code in the algorithm \ref{algo:dgeqr21} that, it contains three steps, 1) computation of a householder vector for each column 2) computation of householder matrix $P$, and 3) update of trailing matrix using $P = I-2vv^T$ where $I$ is an identity matrix. For our experiments, we use Intel C Compiler (ICC) and Intel Fortran Compiler (IFORT). We also use different compiler switches to improve the performance of dgeqr2 in LAPACK on Intel micro-architectures. In Intel Core i7 $4^{th}$ Gen machine which is a Haswell micro-architecture, CPI attained saturates at $1.1$ \cite{tpds1}. In case when compiler switch $-mavx$ is used that enables use of Advanced Vector Extensions (AVX) instructions, the Cycles-Per-Instruction (CPI) attained is increased. This behavior is due to AVX instructions that use Fused Multiply Add (FMA). Due to this fact, the CPI reported by VTune\texttrademark can not be considered as a measure of performance for the algorithms and hence we accordingly double the instruction count reported by Intel VTune\texttrademark.

In case of GPGPUs, dgeqr2 in MAGMA is able to achieve up to 16 Gflops in Tesla C2050 which is 3.1\% of the theoretical peak performance of Tesla C2050 while performance in terms of Gflops/watt is as low as 0.04 Gflops/watt.

\subsection{dgeqrf}
\begin{algorithm}
\caption{dgeqrf (Pseudo Code)}
\label{algo:dgeqrf1}
\begin{algorithmic}[1]
\State Allocate memories for input/output matrices
\For{$i=1$ to $n$}
  \State Compute Householder vectors for block column $m\times k$
  \State Compute $P$ matrix where $P$ is Computed using Householder vectors 
  \State Update trailing matrix using dgemm
\EndFor
\end{algorithmic}
\end{algorithm}

Pseudo code for dgeqrf routine is shown in algorithm \ref{algo:dgeqrf1}. In terms of computations, there is no difference between algorithms \ref{algo:dgeqr21}, and \ref{algo:dgeqrf1}. In a single core implementation, dgeqrf is observed to be 2-3x faster than dgeqr2. The major source of efficiency in dgeqrf is efficient exploitation of processor memory hierarchy and dgemm routine which is a compute bound operation \cite{rob3}\cite{jack1}. In Nvidia Tesla C2050, dgeqrf in MAGMA is able to achieve up to 265 Gflops which is 51.4 \% of theoretical peak performance of Nvidia Tesla C2050 as shown in the figure \ref{fig:cpi_gmm_7} which is 90.5\% of the performance attained by dgemm. In dgeqr2 in MAGMA, performance attained in terms of Gflops/watt is as low as 0.05 Gflops/watt while for dgemm and dgeqrf it is 1.23 Gflops/watt and 1.09 Gflops/watt respectively in Nvidia Tesla C2050 as shown in figure \ref{fig:cpi_gmm_8}. In case of dgqrf in PLASMA, the performance attained is 0.39 Gflops/watt while running dgeqrf in four cores.

\subsection{dgeqr2ht}
\begin{algorithm}
\caption{dgeqr2ht (Pseudo Code)}
\label{algo:dgeqr2ht}
\begin{algorithmic}[1]
\State Allocate memories for input/output matrices
\For{$i=1$ to $n$}
  \State Compute Householder vectors for block column $m\times k$
  \State Compute $PA$ where $PA = A - 2vv^TA$
\EndFor
\end{algorithmic}
\end{algorithm}

Pseudo code for dgeqr2ht is shown in algorithm \ref{algo:dgeqr2ht}. A clear difference between dgeqr2 in the algorithm \ref{algo:dgeqr21}, dgeqrf in the algorithm \ref{algo:dgeqrf1}, and dgeqr2ht in the algorithm \ref{algo:dgeqr2ht} is in updating of trailing matrix. dgeqr2 uses dgemv operation which is a memory bound operation, and dgeqrf uses dgemm operation which is compute bound operation while dgeqr2ht uses an operation that is more dense in-terms of computations resulting in lower $\theta$ where $\theta$ is a rough quantification of parallelism through DAG based analysis of routines dgeqr2, dgeqrf, and dgeqr2ht. In case of no-change in the computations in the improved routine after re-arrangement of computations (in this case fusing of the loops), $\theta$ translates into ratio of number of levels in the DAG of improved routine to number of levels in the DAG of classical routine.
\begin{figure*}%[!h]
\centering
\subfigure[Performance Comparison of PE with Other Platforms for dgemm and dgeqr2ht\label{fig:ht_graph_2}]{\includegraphics[scale = 0.22]{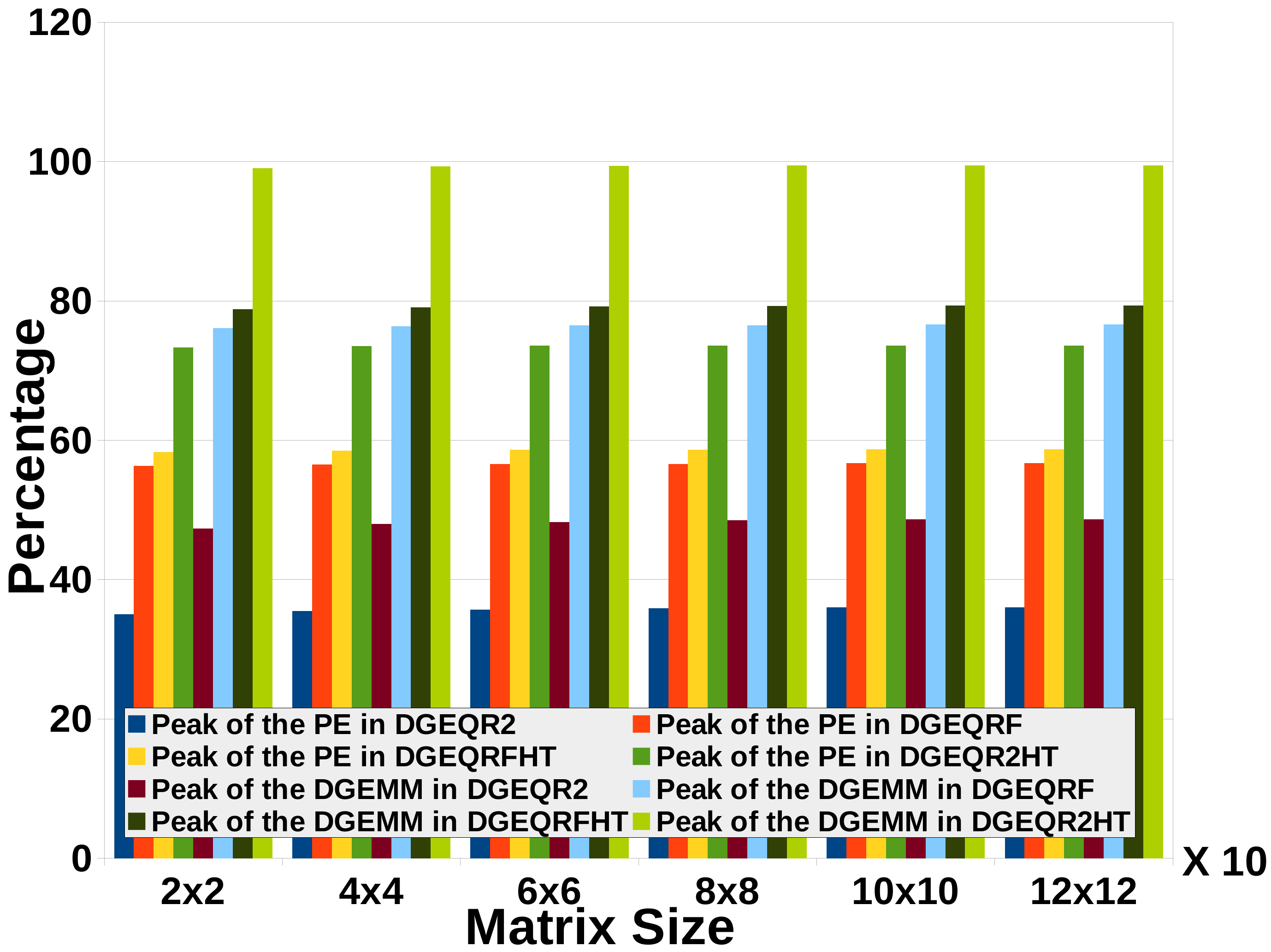}}
\subfigure[Performance of PE In-terms of Theoretical Peak Performance in PE for dgemm \label{fig:ht_graph_3}]{\includegraphics[scale = 0.22]{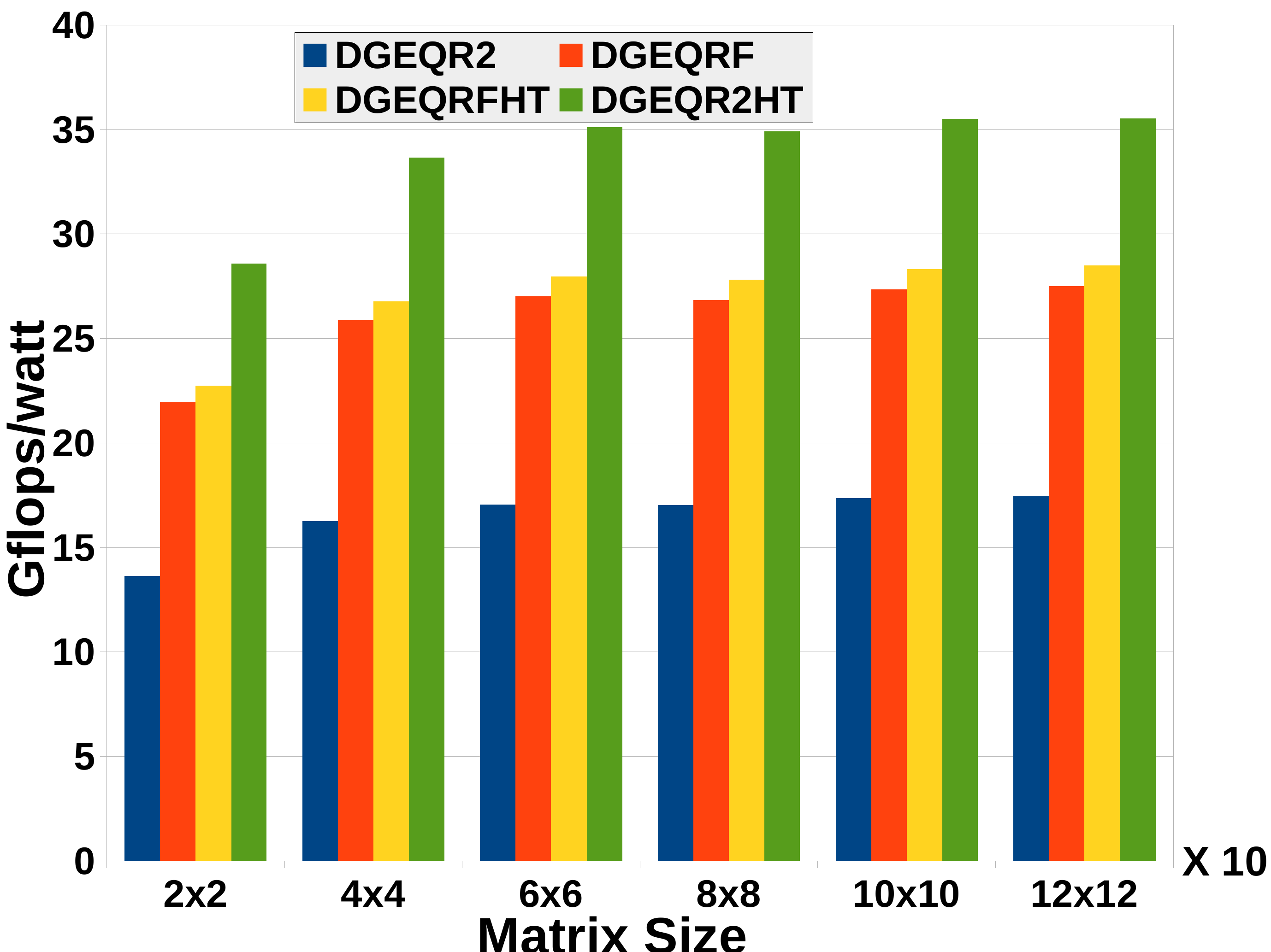}}
\caption{Performance of dgeqr2ht in PE}
\label{fig:ggr_perf}
\end{figure*}
Implementation of dgeqr2ht clearly outperforms implementation of dgeqrf in PE as shown in figure \ref{fig:ht_graph_2}. In the figure \ref{fig:ht_graph_2}, the performance is shown in-terms of percentage of theoretical peak performance of PE and also in-terms of percentage of theoretical peak performance normalized to the performance attained by dgemm in the PE. It can be observed in the figure \ref{fig:ht_graph_2} that the performance attained by dgeqr2ht is 99.3\% of the performance attained by dgemm in the PE. Furthermore, it can be observed in figure \ref{fig:ht_graph_3} that the performance achieved in-terms of Gflops/watt in the PE is 35 Gflops/watt compared to performance attained by dgeqrf which is 25 Gflops/watt. In multicore and GPGPUs, such a fusing does not translate into improvement due to limitations of the platform. Further details of dgeqr2ht implementation can be found in \cite{tpds1}. Based on case studies on dgemm, dgeqr2, dgeqrf, and dgeqr2ht, we make following observations:
\begin{itemize}
	\item dgeqrf in highly tuned software packages like PLASMA and MAGMA attains 16\% and 51\% respectively. This leaves a further scope for improvement in these routines through careful analysis 
	\item Typically, dgeqrf routine that has compute bound operation like dgemm achieves 80-85\% of the theoretical peak performance attained by dgemm in multicore and GPGPUs
	\item dgeqr2ht attains similar performance as dgeqrf in multicore and GPGPUs while it clearly outperforms dgeqrf in PE
\end{itemize}
We further propose generalization in CGR presented in \cite{cgr1} and derive GGR that can outperform dgeqr2ht in PE and in REDEFINE.

%% file: implementation.tex
From equation \ref{eqn:qrd1}, matrix $A_{n\times n}$, annihilation of element in the last row and first column which is (n,1) would require application of one Givens sequence 
\begin{align}\label{eqn:clas_gr1}
	G_{n,1}A = \begin{bmatrix}R^{(1)} \\ 0 \end{bmatrix} 
\end{align}
where matrix $R^{(k)}$ is a matrix with $k$ zero elements in the lower triangle of the matrix and has undergone k-updates. In general Givens matrix is given by equation \ref{eqn:giv_seq}. 
\begin{align}
\label{eqn:giv_seq}
G_{i,j} = diag(I_{i-2}, \tilde G_{i,j}, I_{m-i}) \text{ with } \tilde G = \begin{bmatrix} c & s \\ -s & c \end{bmatrix}  
\end{align} 
where $ c= \frac{A_{i-1,j}}{t}$, $s = \frac{A_{i,j}}{t}$ and $t = \sqrt{A_{i-1,j}^2 + A_{i,j}^2}$. It takes $n-1$ Givens sequences to annihilate $n-1$ elements in the matrix $A$. There is a possibility to apply multiple Givens sequences to annihilate multiple elements in a column of a matrix. Thus, extending equation \ref{eqn:clas_gr1} to annihilate 2-elements in the first column of the matrix $A$
\begin{align}\label{eqn:clas_gr2}
	G_{n-1,1}G_{n,1}A = \begin{bmatrix}R^{(2)} \\ 0\end{bmatrix}
\end{align}
where $(G_{n-1,1}G_{n,1})^T(G_{n-1,1}G_{n,1}) = (G_{n-1,1}G_{n,1})(G_{n-1,1}G_{n,1})^T = I$. Extending further equation \ref{eqn:clas_gr2} to annihilate $n-1$ elements of the first column of the input matrix $A$
\begin{align}\label{eqn:clas_gr3}
	G_{2,1}G_{3,1}...G_{n-1,1}G_{n,1}A = \begin{bmatrix}R^{(n-1)}\\0\end{bmatrix}
\end{align}
where $(G_{2,1}G_{3,1}...G_{n-1,1}G_{n,1})^T(G_{2,1}G_{3,1}...G_{n-1,1}G_{n,1}) $ \\ $= (G_{2,1}G_{3,1}...G_{n-1,1}G_{n,1})(G_{2,1}G_{3,1}...G_{n-1,1}G_{n,1})^T = I$.
Formulation in equation \ref{eqn:clas_gr3} can annihilate $n-1$ elements in the first column of the input matrix $A$. Furthering annihilation of $n-1$ elements to $(n-1) + (n-2)$ elements that results in matrix $R^{(n-1)+(n-2)}$ where $n-1$ elements in the first column and $n-2$ elements in the second column of matrix $R$ are zero as given by equation \ref{eqn:clas_gr4}.
\begin{align}%\label{eqn:clas_gr4}
	(G_{3,2}G_{4,2}...G_{n-1,2}G_{n,2})(G_{2,1}G_{3,1}...G_{n-1,1}G_{n,1})A = \nonumber \\  \begin{bmatrix}R^{(n-1)+(n-2)}\\0\end{bmatrix} \label{eqn:clas_gr4}
\end{align}

where $((G_{3,2}G_{4,2}...G_{n-1,2}G_{n,2})(G_{2,1}G_{3,1}...G_{n-1,1}G_{n,1}))$ \\ $((G_{3,2}G_{4,2}...G_{n-1,2}G_{n,2})(G_{2,1}G_{3,1}...G_{n-1,1}G_{n,1}))^T =$\\$ ((G_{3,2}G_{4,2}...G_{n-1,2}G_{n,2})(G_{2,1}G_{3,1}...G_{n-1,1}G_{n,1}))^T$ \\ $((G_{3,2}G_{4,2}...G_{n-1,2}G_{n,2})(G_{2,1}G_{3,1}...G_{n-1,1}G_{n,1})) = I$. To annihilate $\frac{n(n-1)}{2}$ elements in the lower triangle of the input matrix $A$, it takes $n-1$ sequences and the Givens matrix shrinks by one row and one column with each column annihilation. Further generalizing equation \ref{eqn:clas_gr4},

\begin{figure}[!ht]
	\centering
	\includegraphics[scale = 0.30]{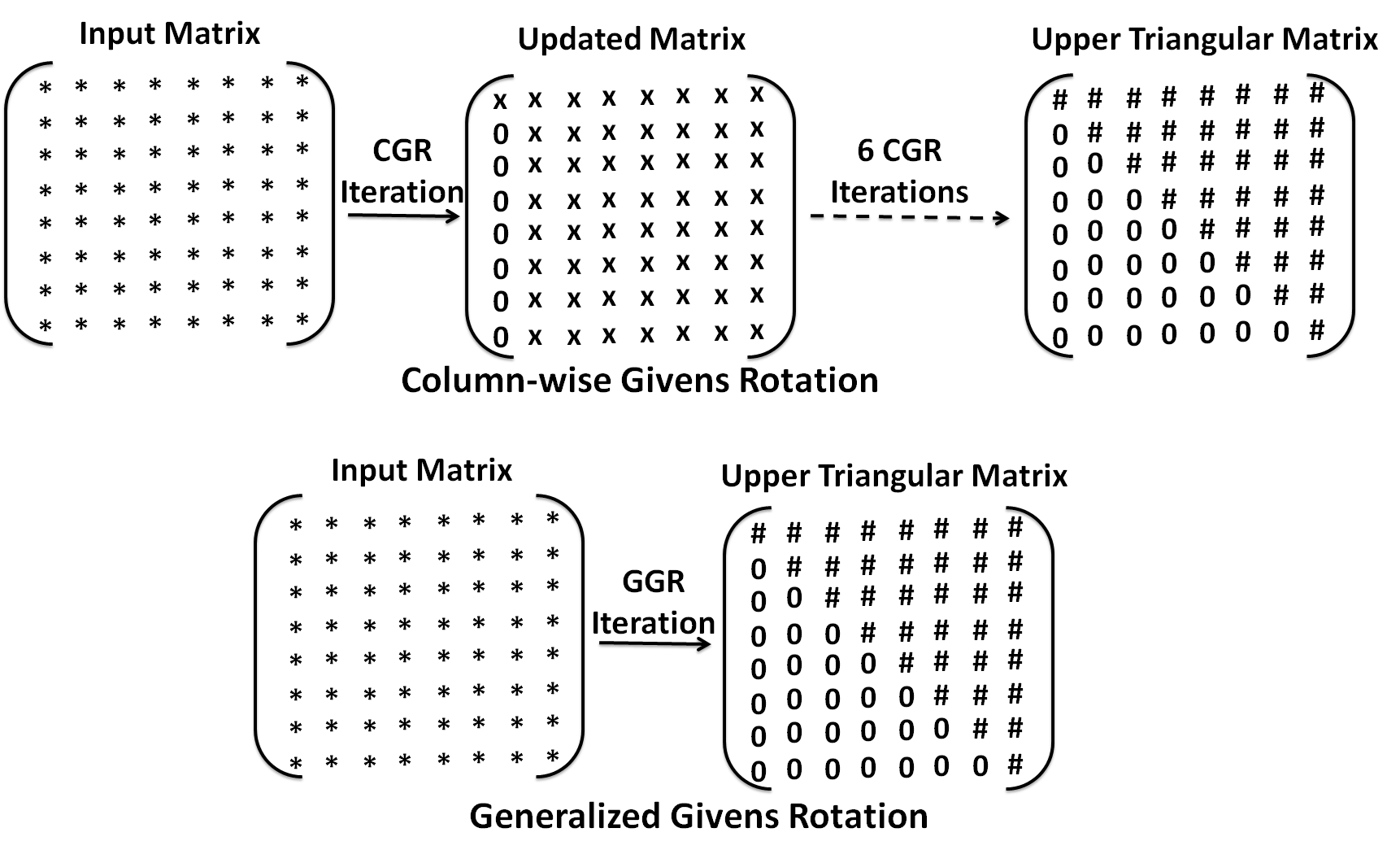}
	\caption{CGR and GGR}
	\label{fig:ggr1}
\end{figure}

\begin{align}\label{eqn:ggr1}
& (G_{n,n-1})(G_{n-1,n-2}G_{n,n-2})....(G_{3,2}G_{4,2}...G_{n-1,2}G_{n,2}) \nonumber \\ &(G_{2,1}G_{3,1}...G_{n-1,1}G_{n,1})A = \begin{bmatrix}R^{(n-1)+(n-2)}\\0\end{bmatrix}
\end{align}

\noindent and $(G_{n,n-1})(G_{n-1,n-2}G_{n,n-2})....(G_{3,2}G_{4,2}...\\ G_{n-1,2}G_{n,2})(G_{2,1}G_{3,1}...G_{n-1,1}G_{n,1}) = Q^T$ and $QQ^T = Q^TQ = I$. Equation \ref{eqn:ggr1} represents GGR in equation form. 
 
%\begin{align}
%	G_{2,1}G_{3,1}...G_{n-1,1}G_{n,1}A = \begin{bmatrix}R^{(n-1)}\\0
%\end{align}
\begin{algorithm}
\caption{Generalized Givens Rotation  (Pseudo code)}
\label{algo:dgeqrggr1}
\begin{algorithmic}
\State{Allocate memory for input/output matrices and vectors}
\For{$i=1$ to $n$}
\State{Compute $2-norm$ of the column vector}
\State{Update row $1$}
\State{update rows $2$ to $n$}
\EndFor
\end{algorithmic}
\end{algorithm}

A pictorial view for $8\times 8$ matrix that compares CGR with GGR is shown in figure \ref{fig:ggr1}, and GGR pseudo code is given in algorithm \ref{algo:dgeqrggr1}. It can be observed in the figure \ref{fig:ggr1} that CGR operates column-wise and takes total $7$ iterations to upper traingularize the matrix of size $8\times 8$ while GGR operates column-wise as well as row-wise and can upper triangularize matrix in a single iteration. It can be observed in the algorithm \ref{algo:dgeqrggr1} that the update of the first row and the rest of the rows can be executed in parallel. Furthermore, there also exist parallelism across the out loop iterations in GGR. Theoretically, classical GR takes $\frac{n(n-1)}{2}$ iterations to upper triangularize an input matrix of size $n\times n$, and CGR takes $n-1$ iterations to upper triangularize an input matrix of size $n\times n$ while GGR can upper triangularize a matrix of size $n\times n$ in $1$ iteration. 

However, in practical scenario, it is not possible to accommodate large matrices in the registers or Level 1 (L1) cache memory. Hence, a sophisticated matrix partitioning schemes are required to efficiently exploit the memory hierarchy in multicores and GPGPUs. 

\subsection{GGR in Multicore and GPGPU}
For multiocre realization of GGR, we use PLASMA framework and for GPGPU realization, we use MAGMA framework depicted in figure \ref{fig:plasma_magma}.
\subsubsection{GGR in PLASMA}
To implement GGR in multicore architectures, we first implement dgeqr2ggr routine in LAPACK and we use that routine in PLASMA. GGR implementation in LAPACK is shown in algorithm \ref{algo:dgeqrggr2} as a pseudo code. It can be observed in the algorithm \ref{algo:dgeqrggr2} that the most computationally intensive part in GGR is $update$ function. In our implementation, $update$ function becomes part of BLAS while in LAPACK, we implement $dgeqr2ggr$ function that is wrapper function of $update$ function that calls $update$ function $n$ times for input matrix of size $n\times n$ as shown in algorithm \ref{algo:dgeqrggr2}. 

\begin{algorithm}
\caption{lapack\_dgeqr2ggr (GGR in LAPACK)  (Pseudo code)}
\label{algo:dgeqrggr2}
\begin{algorithmic}
\State{Allocate memory for input/output matrices and vectors}
\For{$i=1$ to $n$}
\State{update(L, A(i,i), A, LDA, I, N, M, Tau(i), Beta)}
\EndFor
\end{algorithmic}
\begin{algorithmic}
\State{update()}
\State{Initialize $k$,$l$, and $s$ vectors to $0$}
\State{Calculate 2-norm of the column vector}
\State{Calculate $k$, $l$, and $s$ vectors}
\State{Update row $i$ of the matrix}
\State{Update row $i+1$ to $n$ using $k$, $l$, and $s$ vectors}
\end{algorithmic}
\end{algorithm}

%\begin{figure*}%[!h]
%\centering
%\subfigure[Speed-up in dgeqrggr over dgeqr2, dgeqrf, dgeqrfht, and dgeqr2ht in PE\label{fig:ggr_graph_14}]{\includegraphics[scale = 0.23]{figures/ggr_graph_1}}
%\subfigure[Performance of dgeqrggr, dgeqr2, dgeqrf, dgeqrfht, and dgeqr2ht in PE In-terms of Theoretical Peak Performance Normalized to Performance of dgemm in the PE\label{fig:ggr_graph_15}]{\includegraphics[scale = 0.23]{figures/ggr_graph_2}}
%\caption{Performance of dgeqrggr in PE}
%\label{fig:dgeqrggr_performance_sot}
%\end{figure*}

Performance comparison of dgeqr2ggr, dgeqrfggr, dgeqr2, dgeqrf, dgeqr2ht, and dgeqrfht in LAPACK and PLASMA is shown in figure \ref{fig:ggr_perf_cpu_gpu}. It can be observed in the figure \ref{fig:ggr_perf_cpu_gpu} that despite more computations in dgeqr2ggr, the performance of dgeqr2ggr is comparable to dgeqr2, and performance of dgeqrfggr is comparable to dgeqrf in LAPACK and PLASMA. We compare run-time of the different routines normalized to dgemm performance in respective software packages since total number of computations in HT, MHT, and GGR are different. Furthermore, in our implementation of GGR in PLASMA, we use dgemm for updating trailing matrix.

\subsubsection{GGR in MAGMA}
\begin{figure}[!ht]
	\centering
	\includegraphics[scale = 0.32]{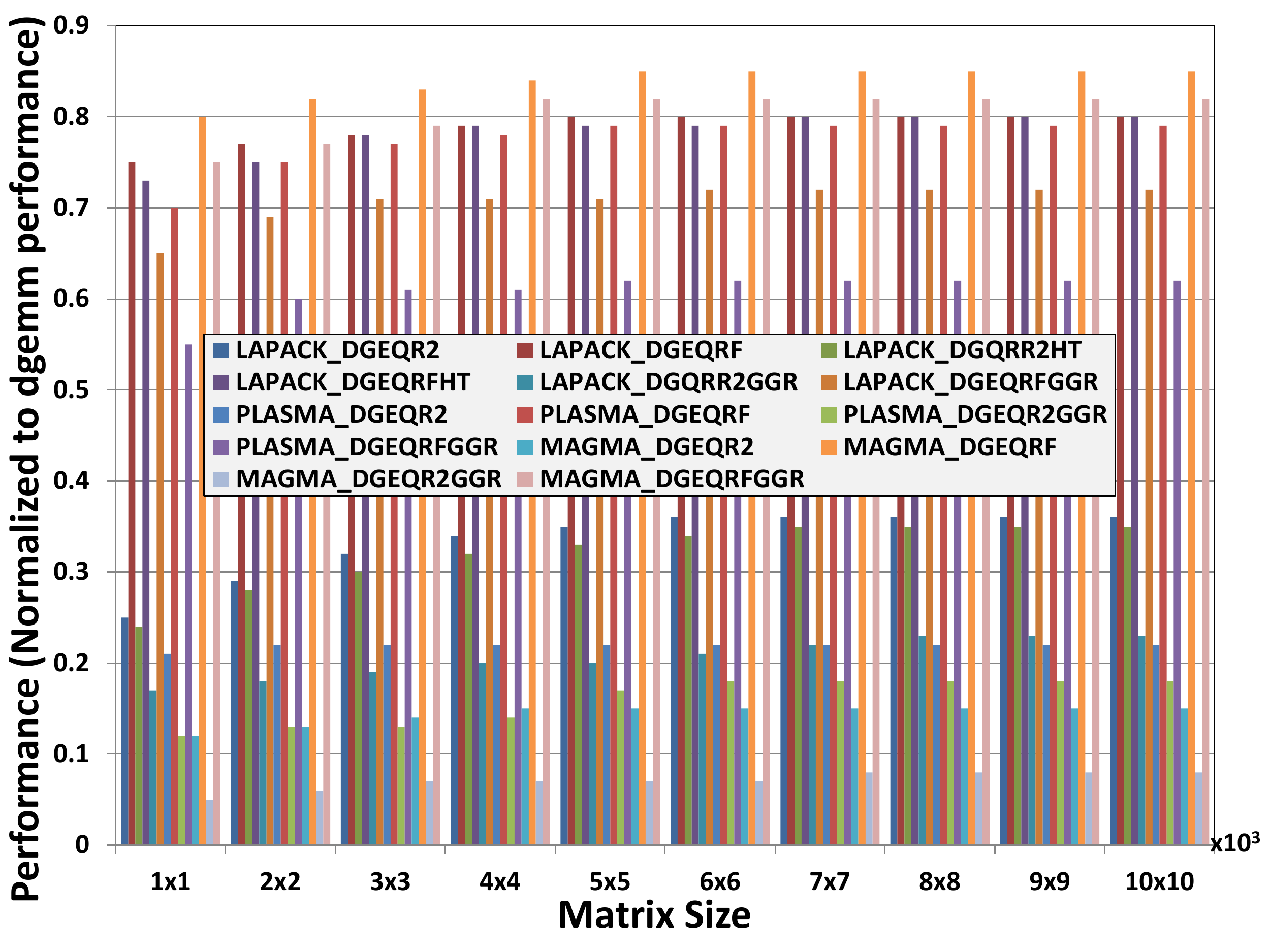}
	\caption{Performance of GGR in Different Packages and Platforms}
	\label{fig:ggr_perf_cpu_gpu}
\end{figure}

For implementation of magma\_dgeqr2ggr, we insert routines in MAGMA\_BLAS. Pseudo code for the inserted routine is shown in algorithm \ref{algo:dgeqrggr3}.
\begin{algorithm}
\caption{magma\_dgeqr2ggr (GGR in MAGMA)  (Pseudo code)}
\label{algo:dgeqrggr3}
\begin{algorithmic}
\State {dnrm2$<<<$grid, threads,0,queue$->$cuda\_stream()\par $>>>$(n,m,dC,lddc,dv,ddot)}
\State{klvec$<<<$grid,threads,0,queue$->$cuda\_stream() \par $>>>$(n,m,ddot,lddc,dv,dk,dl)}
\State{dtmup$<<<$n,threads,0,queue$->$cuda\_stream() \par $>>>$(n,m,dC,lddc,ddot,dv,dk,dl)}
\end{algorithmic}
\begin{algorithmic}
\State{dnrm2()}
\For{$k=1$ to $m$}
\For{$j=0$ to $BLOCK\_SIZE$}
	\State{$w += A[ldda\ast (row)+(m-1-j-k)]\ast V\_shared[j]$}
	\If{$row == 0$}
		\State{$dot[ldda\ast (row) + (m-1-k-j)] = -copysign(sqrt(w),vector[0])$}
		\ElsIf{$row!=0$}
		\State{$dot[ldda\ast (row) + (m-1-k-j)] = w$}
	\EndIf
\EndFor
\EndFor
\State{klvec()}
\If{$row == 0$}
	\State{dl[row] = 0}
	\State{dk[row] = 1/dot[0]}
\ElsIf{$row == m-1$}
	\State{$dl[row] = dv[row]/dot[row-1]$}
	\State{$dk[row] = dv[row-1]/dot[row-1]$}
\ElsIf{$(row > 0) \&\& (row < (m-1))$}
	\State{$dl[row]  = dot[row]/dot[row-1]$}
	\State{$dk[row] = dv[row-1]/(dot[row]\ast dot[row-1])$}
\EndIf
\State{dtmup()}
\State{tx = threadIdx.x}
\State{$dC[m-1] = (dk[m-1]\ast dC[m-1]) - (dl[m-1]\ast dC[m-2])$}
\For{$j = m-2-tx$ to $1$} 
\State{$dC[j] = (dk[j]\ast dot[j]) - (dl[j]\ast dC[j-1])$}
\State{$j = j-BLOCK\_SIZE$}
\EndFor
\State{$dC[0] = dk[0]\ast dot[0]$}
\end{algorithmic}
\end{algorithm}
The implementation consists of $3$ functions, $drnm2$ that computes 2-norm of the column vector, computation of $k$, and $l$ vectors by function $klvec$ and update of trailing matrix by $dtmup$ function. It can be observed in the algorithm \ref{algo:dgeqrggr3} that the most computationally intensive part in the routine is $dtmup$ function. There are two routines implemented, 1) dgeqr2ggr where trailing matrix is updated using method shown in equation \ref{eqn:cgr_4}, 2) dgeqrfggr where trailing matrix is updated using dgemm. Performance of dgeqr2ggr and dgeqrfggr is shown in figure \ref{fig:ggr_perf_cpu_gpu}. It can be observed in the figure \ref{fig:ggr_perf_cpu_gpu} that the performance of dgeqr2ggr is similar to performance of dgeqr2 in MAGMA while performance of dgeqrfggr is similar to the performance of dgeqrf in magma. Despite abundant parallelism available in dgeqr2ggr, GPGPUs are not capable of exploiting this parallelism due to serialization of the routine while in dgeqrfggr, GPGPUs perform similar to dgeqrf due to dominance of dgemm in the routine. 
\subsection{GGR in PE}
\begin{figure*}[!ht]
	\centering
	\includegraphics[scale = 0.25]{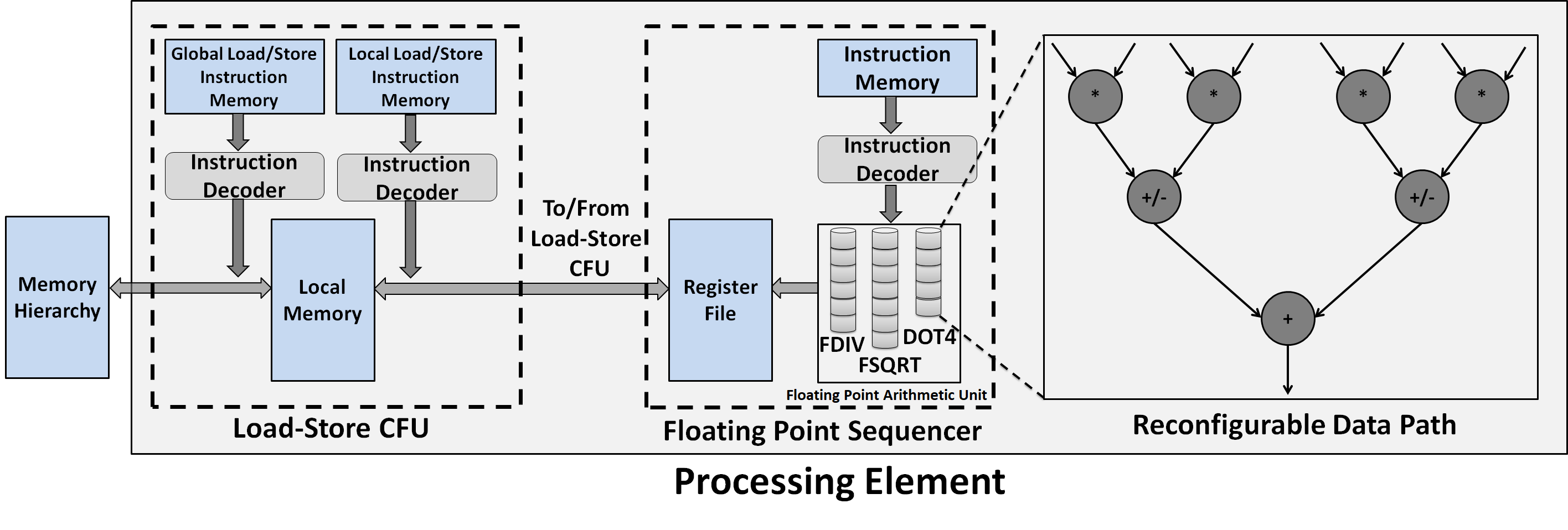}
	\caption{Processing Element with RDP}
	\label{fig:cfuredefine1}
\end{figure*}

For implementation of dgeqr2ggr, and dgeqrfggr in PE, we use similar method as presented in \cite{tpds1}. We perform DAG based analysis of GGR and identify macro operations in the DAGs. These macro operations are then realized on RDP that is tightly coupled to PE as shown in figure \ref{fig:cfuredefine1}. PE consists of two modules, 1) Floating Point Sequencer (FPS), and 2) Load-Store CFU.   

\begin{figure}[!ht]
	\centering
	\includegraphics[scale = 0.25]{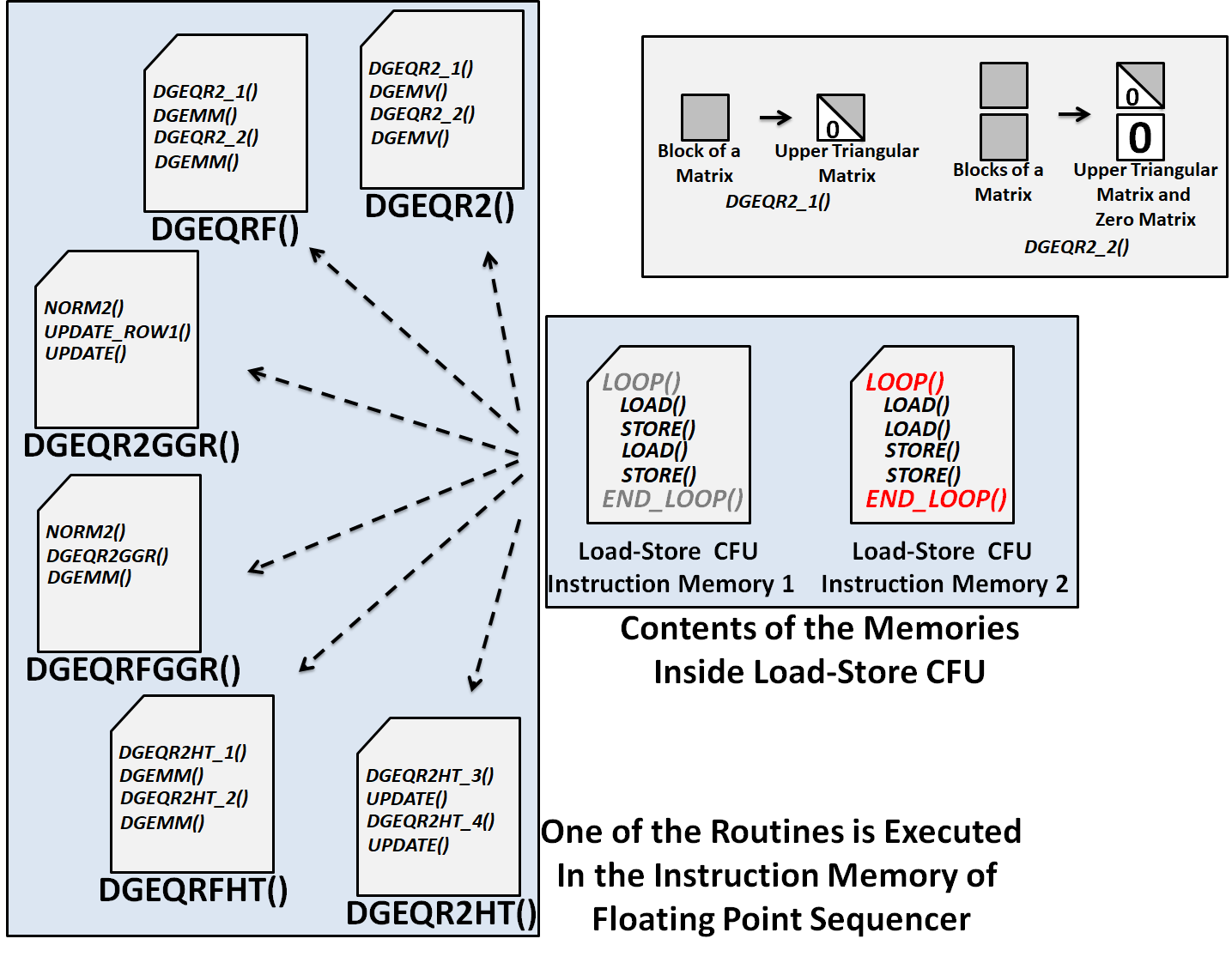}
	\caption{Implementation of dgeqr2, dgeqrf, dgeqrfht, dgeqr2ht, dgeqr2ggr, and dgeqrfggr in PE}
	\label{fig:pe_perf}
\end{figure}
\begin{figure}[!ht]
	\centering
	\includegraphics[scale = 0.20]{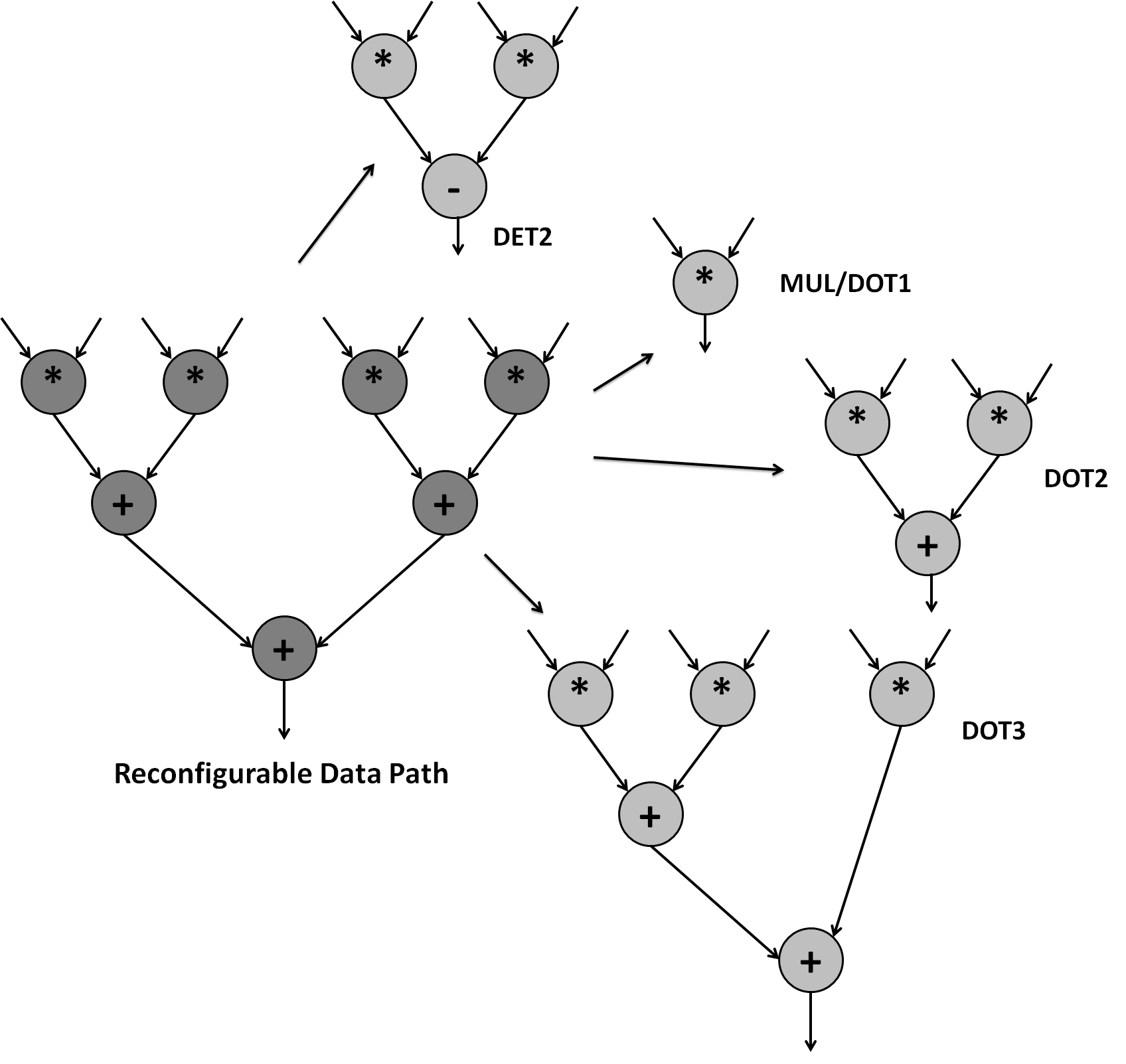}
	\caption{Different Configurations in RDP to Support Implementation of dgeqr2ggr}
	\label{fig:rdp_update}
\end{figure}

FPS performs double precision floating point computations while Load-Store CFU is responsible for loading and storing of data in registers, and Local Memory (LM) to/from the Global Memory (GM). Operation in the PE can be defined by following steps:
\begin{itemize}
	\item Send a load request to GM for input matrix and store input matrix elements in LM
	\item Move input matrix elements to registers in the FPS
	\item Perform computations and store the results back to LM
	\item Write results back to GM if the computed elements are the final output or if they are not to be used immediately
\end{itemize}
Up to 90\% of overlap in computation and communication is attained in the PE for dgemm as presented in \cite{Merc1}. To identify macro operations, considering example of $4\times 4$ matrix shown in the figure \ref{fig:fig2} and equation \ref{eqn:cgr_4}. It can be observed in the figure \ref{fig:fig2} that computing Givens Generation (GG) is an square root of inner product while computing update of the first row is similar to inner product. Furthermore, computing rows $2$,$3$, and $4$ is determinant of $2\times 2$ matrix. We map these row updates on RDP as shown in figure \ref{fig:rdp_update}. It can be observed in the figure \ref{fig:rdp_update} that the RDP can be re-morphed to perform scalar multiplication (MUL/DOT1), inner product of 2-element vectors (DOT2), inner product of 3-element vectors, determinant of $2\times 2$ matrix (DET2), and innter product of 4-element vectors (DOT4) operations. In our implementation we ensure that the reconfiguration of RDP is minimal to improve energy efficiency of the PE. We introduce custom instructions like DOT4, DOT3, DOT2, DOT1, and DET2 instructions in PE to implement these macro operations in RDP along with instruction that can reconfigure RDP. Different routines are realized using these instructions as shown in figure \ref{fig:pe_perf}. In the figure \ref{fig:pe_perf}, it can be observed that irrespective of routine for QR factorization implemented in PE, the communication pattern remains consistent across the routines. In our implementation of dgeqr2ggr, we ensure that RDP is configured to perform two DET2 instructions in parallel that maximizes resource utilization of RDP. In our implementation of dgeqr2ggr, instructions in the two function UPDATE\_ROW1 and UPDATE are merged such that the pipeline stalls in the RDP are minimized. Such an approach is not possible in implementation of dgeqrfggr or dgeqrfht since trailing matrix is updated using dgemm routine and until the matrix required to update the trailing matrix update is not computed, trailing matrix update can not be processed.  
\begin{figure*}%[!h]
\centering
\subfigure[Speed-up in dgeqr2ggr over dgeqr2, dgeqrf, dgeqrfht, and dgeqr2ht in PE\label{fig:ggr_graph_1}]{\includegraphics[scale = 0.25]{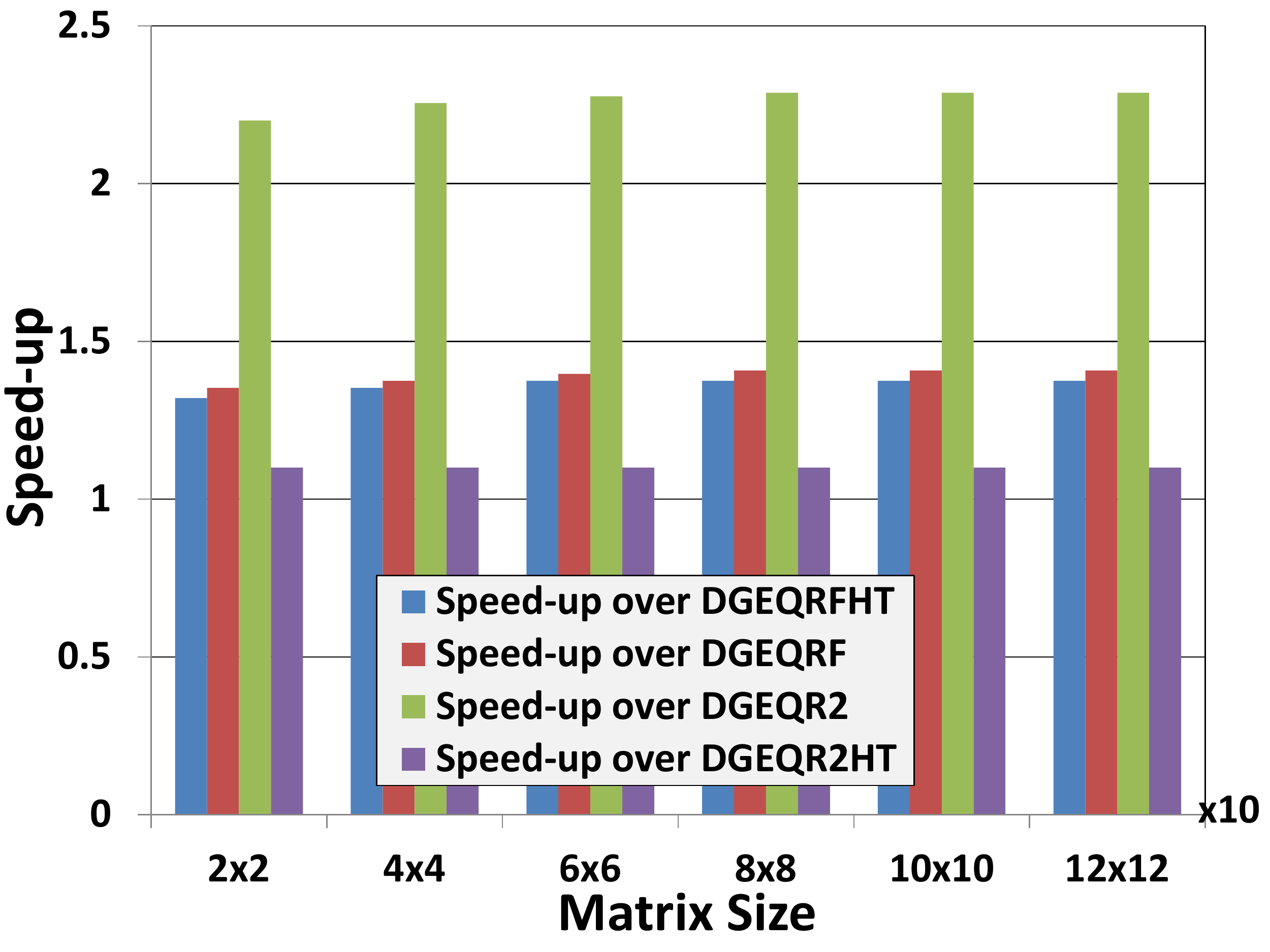}}
\subfigure[Performance of dgeqr2ggr, dgeqrfggr, dgeqr2, dgeqrf, dgeqrfht, and dgeqr2ht in PE In-terms of Theoretical Peak Performance Normalized to Performance of dgemm in the PE\label{fig:ggr_graph_2}]{\includegraphics[scale = 0.25]{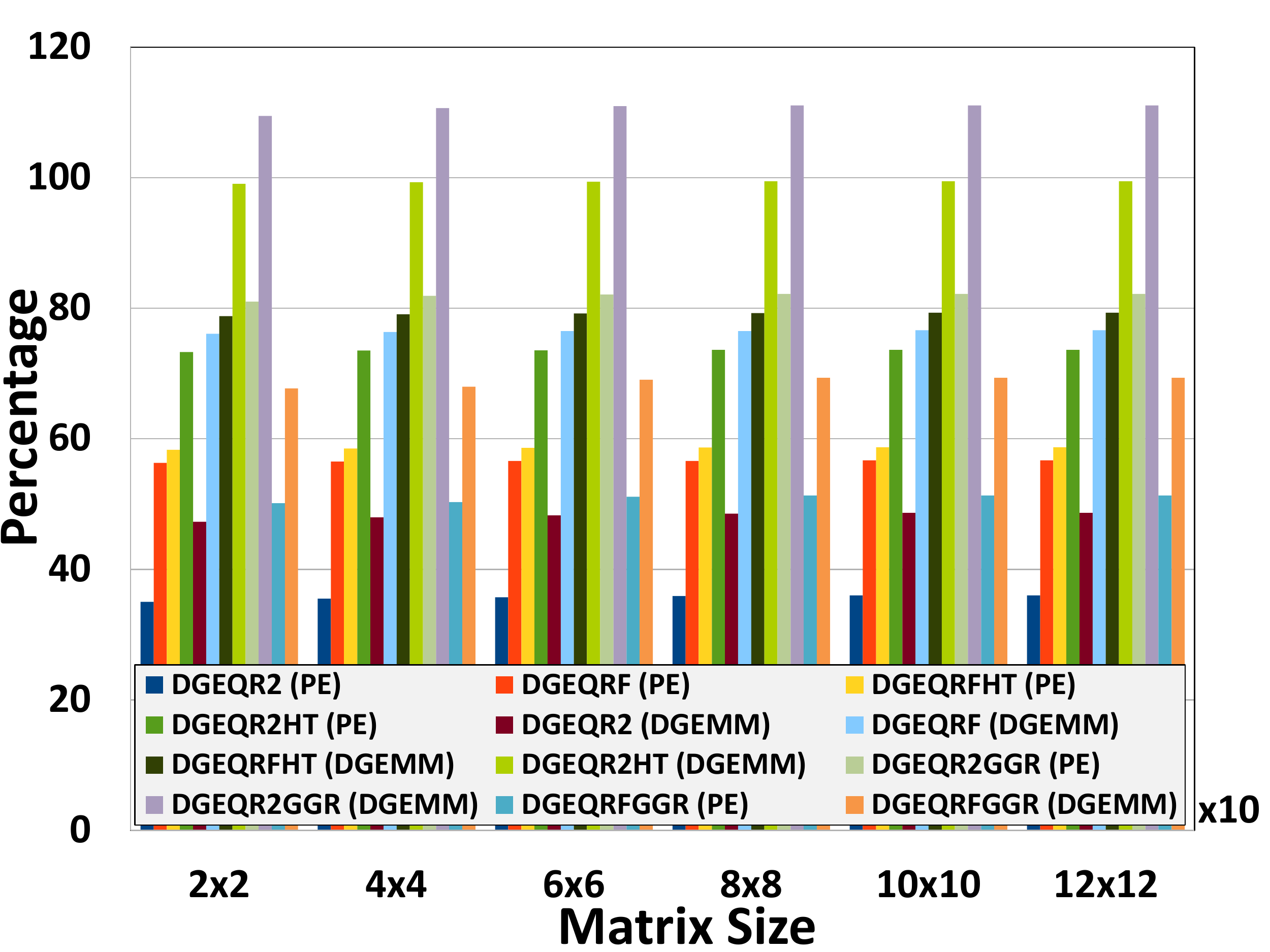}}
\subfigure[Performance of dgeqr2ggr, dgeqr2, dgeqrf, dgeqrfht, and dgeqr2ht in PE In-terms of Gflops/watt\label{fig:ggr_graph_3}]{\includegraphics[scale = 0.25]{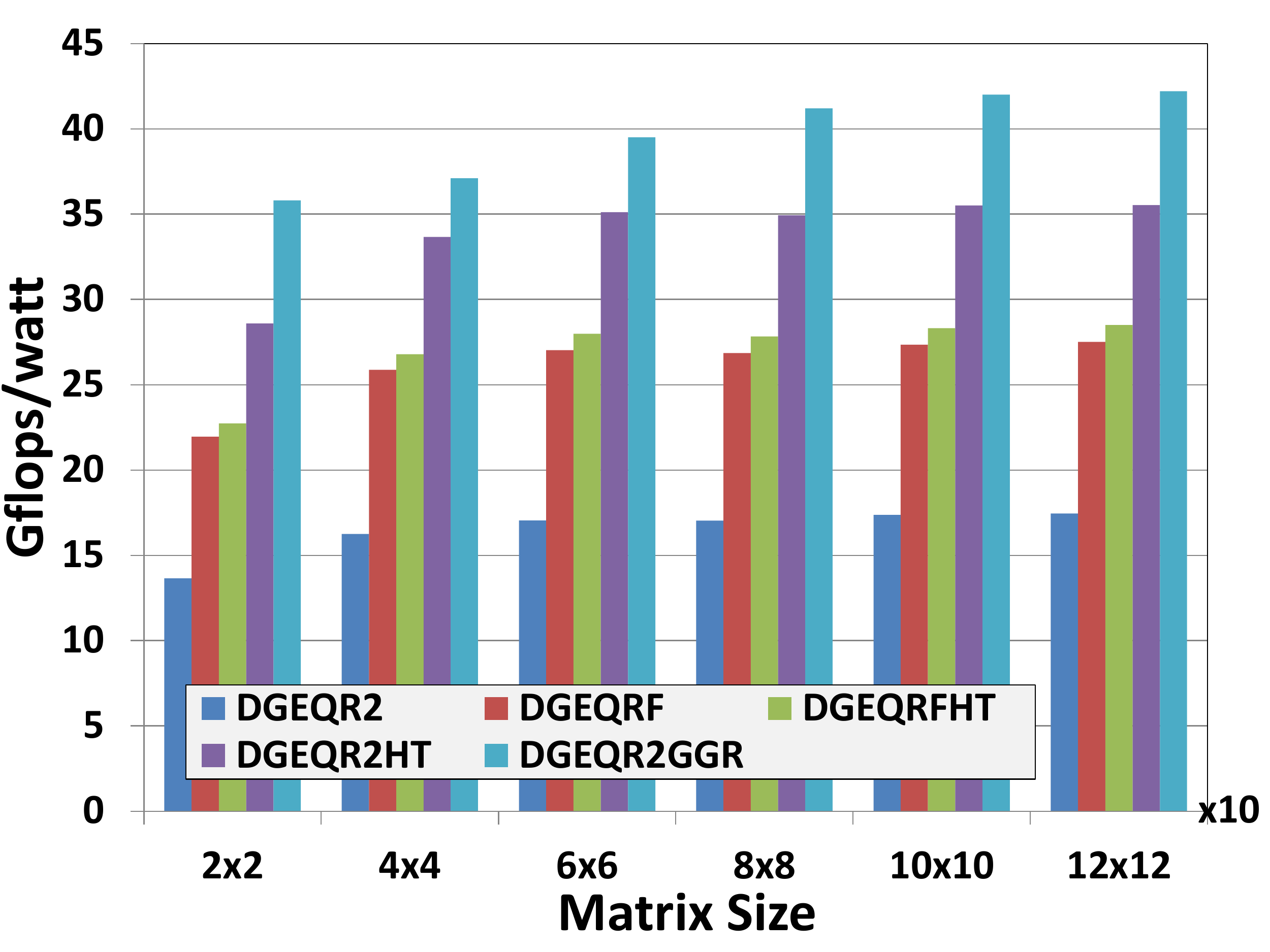}}
\subfigure[Performance Improvement of dgeqr2ggr, dgeqr2, dgeqrf, dgeqrfht, and dgeqr2ht in PE over Different Platforms\label{fig:ggr_graph_4}]{\includegraphics[scale = 0.25]{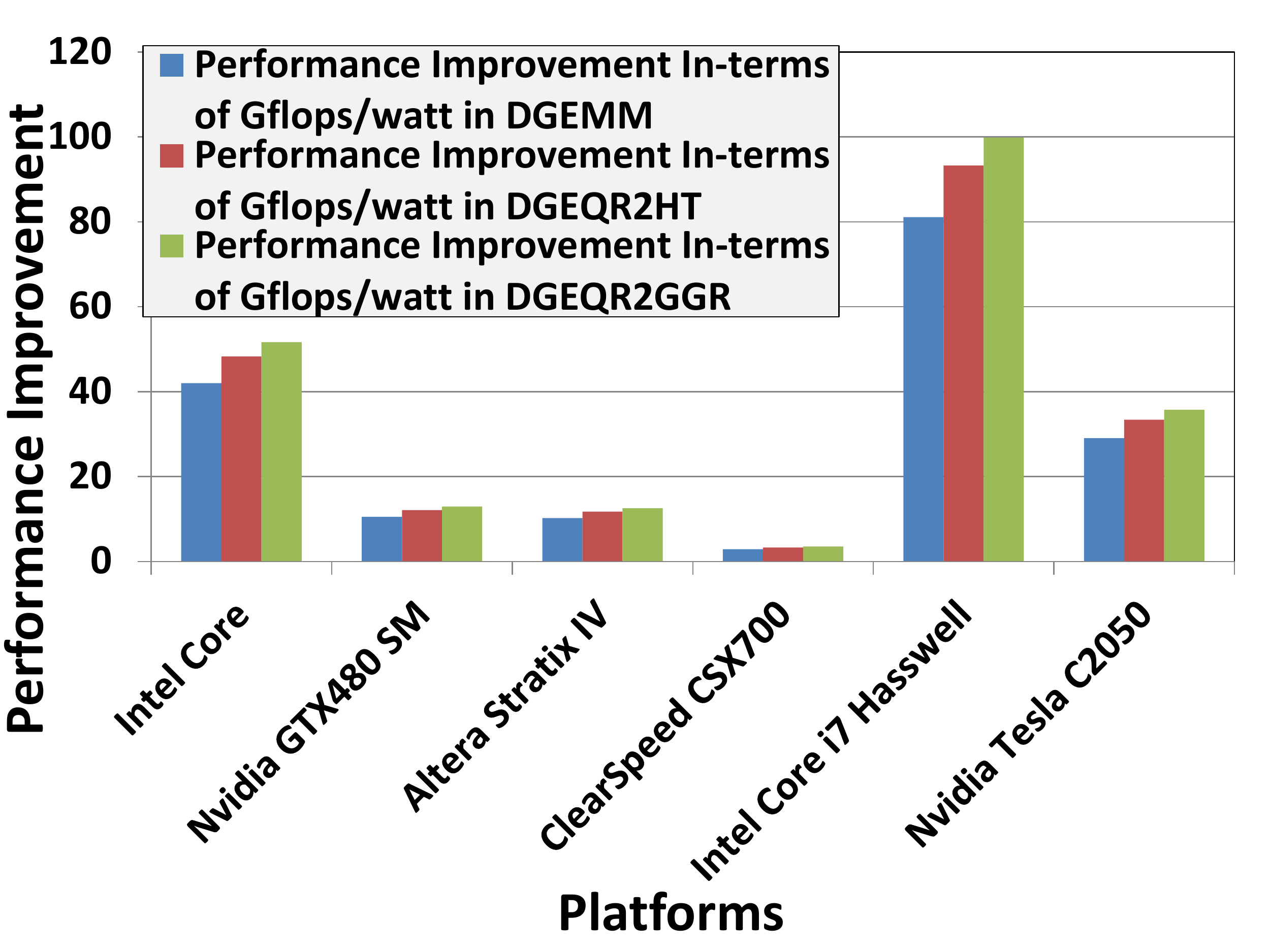}}
\caption{Performance of dgeqr2ggr in PE}
\label{fig:dgeqrggr_performance}
\end{figure*}

Speed-up in dgeqr2ggr over different routines is shown in figure \ref{fig:ggr_graph_1}. It can be observed in the figure \ref{fig:ggr_graph_1} that the speed-up attained in dgeqr2ggr over other routines range between 1.1 to 2.25x. Performance in-terms of the percentage of peak performance normalized to the performance attained by dgemm for different routines for QR factorization is shown in figure \ref{fig:ggr_graph_2}. A counter-intuitive observation that can be made here is that dgeqr2ggr can achieve performance that is higher than the performance attained by dgemm in the PE while dgeqr2ht performance reported in \cite{tpds1} is same as performance attained by dgemm. dgeqr2ggr can achieve performance that is up to 82\% of the theoretical peak of PE. dgeqr2ggr attains 10\% higher Gflops/watt over dgeqr2ht which is the best performing routine as reported in \cite{tpds1} and \cite{Merc2}. Furthermore, improvement in dgeqr2ggr over other platforms is shown in figure \ref{fig:ggr_graph_4}.  It can be observed in the figure \ref{fig:ggr_graph_4} that the performance improvement in PE for dgeqr2ggr over dgeqr2ggr in off-the-shelf platforms is ranging from 3-100x. 

%% file: parallel_imp.tex
\begin{figure*}[!ht]
	\centering
	\includegraphics[scale = 0.15]{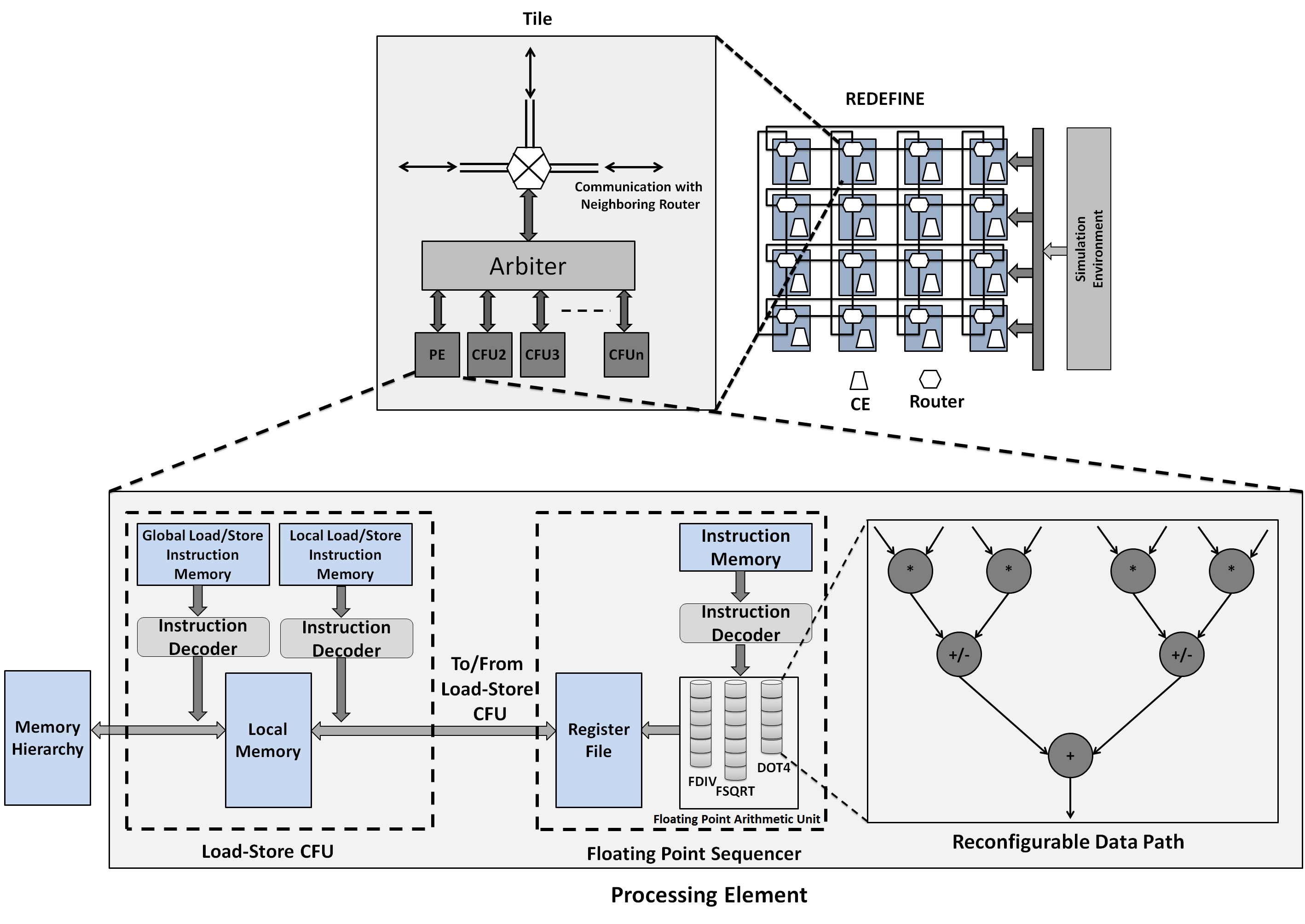}
	\caption{Experimental Setup in REDEFINE with PE as a CFU}
	\label{fig:redefine1}
\end{figure*}

An experimental setup for implementation of dgeqrfggr and dgeqr2ggr is shown in figure \ref{fig:redefine1} where PE that outperforms other off-the-shelf platforms for DLA computations is attached as a CFU to the Routers in REDEFINE CEs. 

\begin{figure}[!ht]
	\centering
	\includegraphics[scale = 0.35]{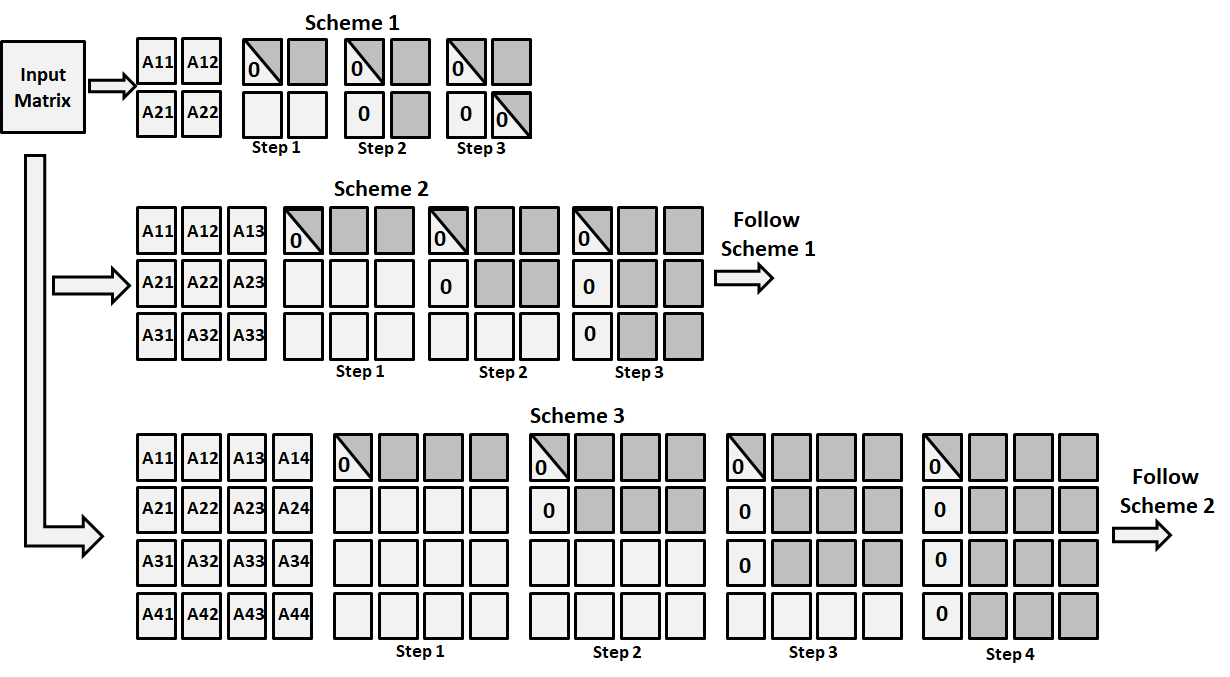}
	\caption{Matrix Partitioning and Scheduling on REDEFINE Tile Array}
	\label{fig:redefine2}
\end{figure}

\begin{figure*}%[!h]
\centering
\subfigure[Speed-up in dgeqr2ggr, dgeqrfggr, dgeqr2, dgeqrf, dgeqr2ht, dgeqrfht Routines in REDEFINE over Their Sequential Realization in PE\label{fig:ht_graph_5}]{\includegraphics[scale = 0.25]{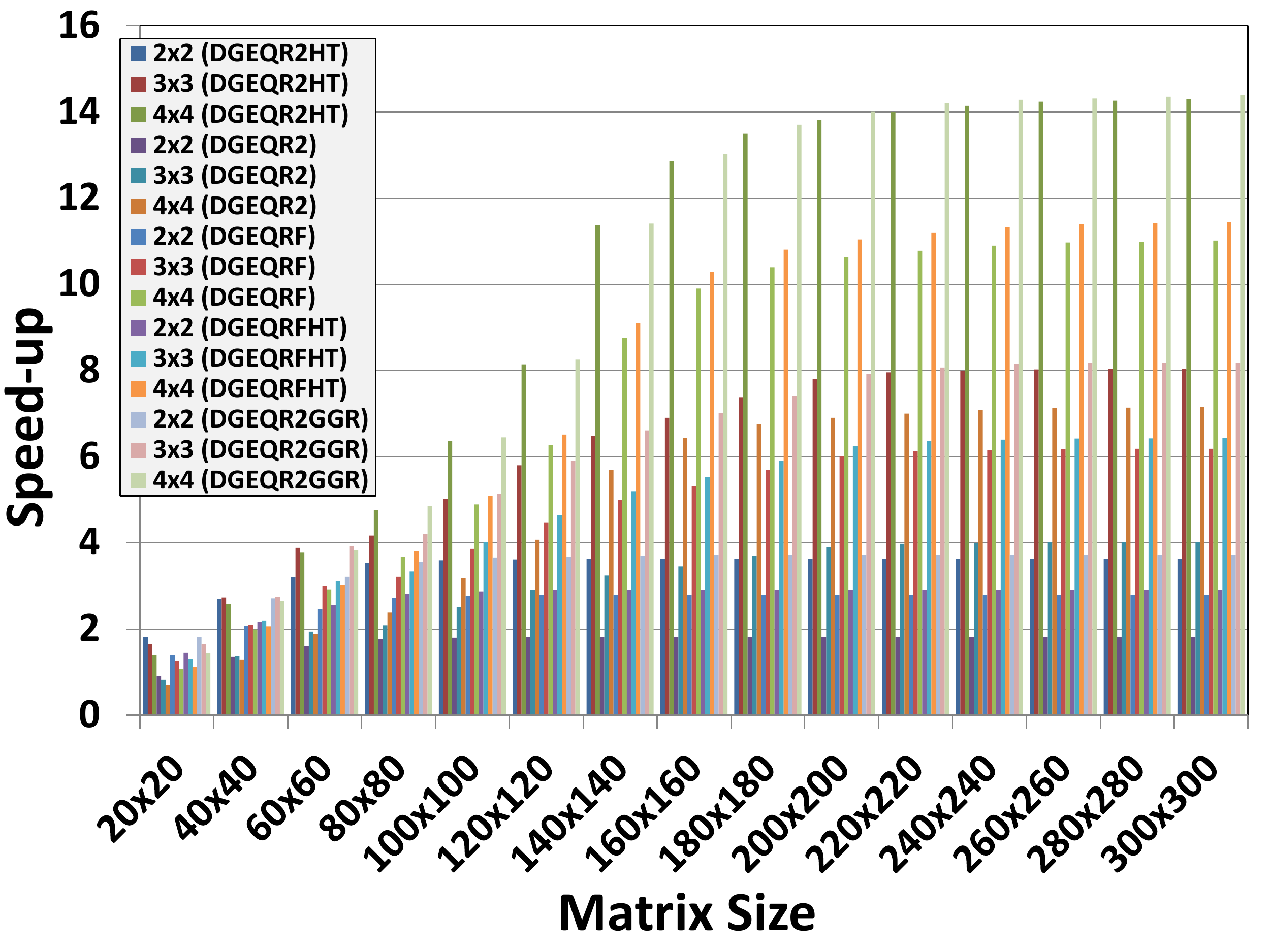}}
\subfigure[Performance of dgeqr2ggr, dgeqrfggr, dgeqr2, dgeqrf, dgeqrfht, and dgeqr2ht in REDEFINE In-terms of Theoretical Peak Performance of REDEFINE\label{fig:ht_graph_6}]{\includegraphics[scale = 0.25]{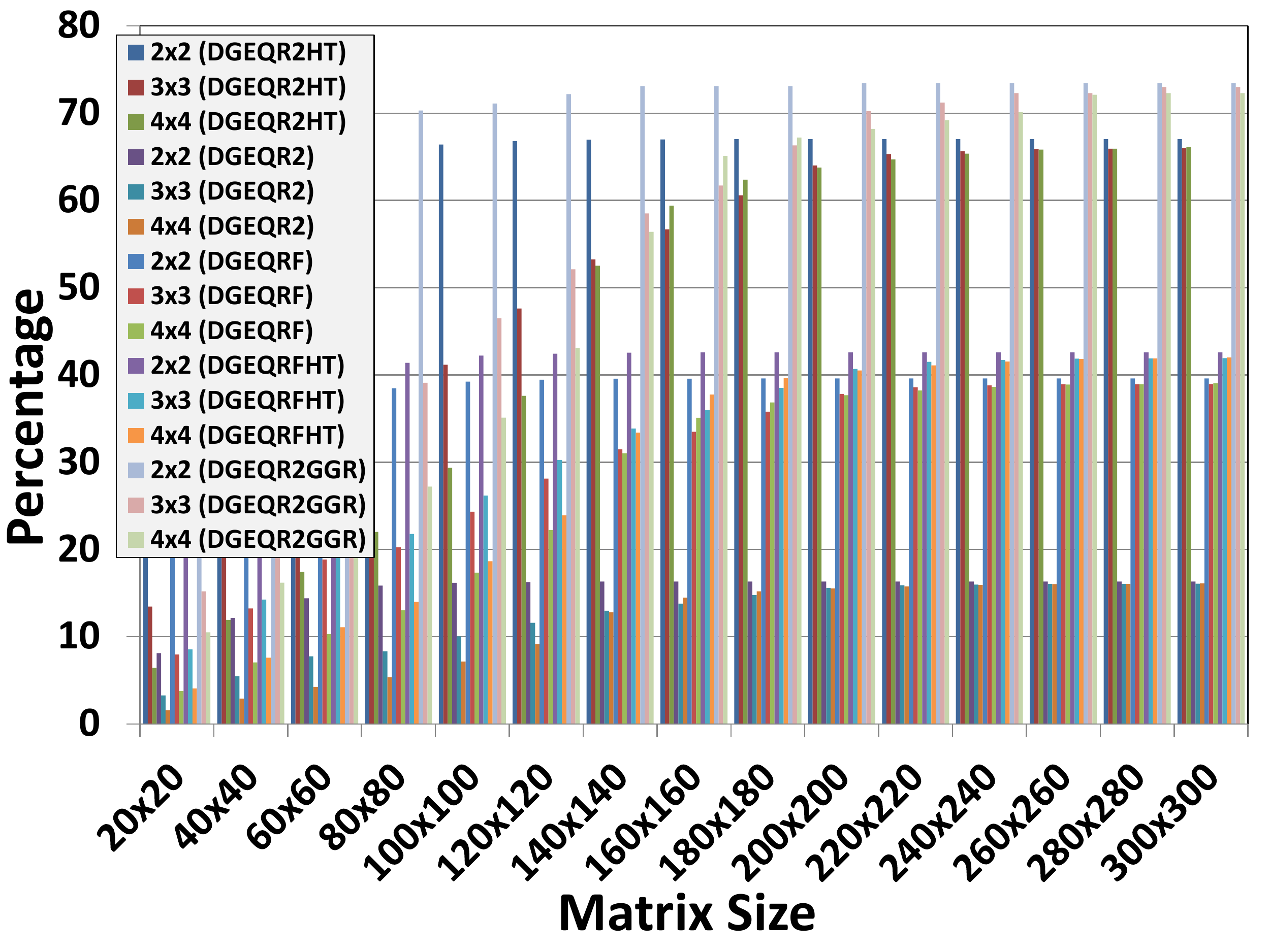}}
\caption{Performance of dgeqr2ggr in PE}
\label{fig:parallel_perf}
\end{figure*}
For implementation of dgeqrfggr and dgeqr2ggr in REDEFINE, it requires an efficient partitioning and mapping scheme that can sustain computation to communication ratio that is commensurate with the hardware resources of the platform, and also ensures scalability. We follow similar strategy presented in \cite{tpds1} for realization of dgeqrfggr and dgeqr2ggr in REDEFINE and propose a general scheme that is applicable for the input matrix of any size. For our experiments, we consider $3$ different configurations in REDEFINE consisting of $2\times 2$, $3\times 3$, and $4\times 4$ Tile arrays. Assuming that input matrix is of the size $N\times N$ and REDEFINE Tile array is of size $K\times K$, the input matrix can be partitioned into the blocks of $\frac{N}{K}\times \frac{N}{K}$ sub-matrices. Since, objective of our experiment is to show scalability of our solution, we choose $N$ and $K$ such that $N\%K=0$. Matrix partitioning and REDEFINE mapping is depicted in figure \ref{fig:redefine2}. As shown in the figure \ref{fig:redefine2}, for Tile array of size $2\times 2$, we follow scheme $1$ where input matrix is partitioned in to sub-matrices of size $2\times 2$. For Tile array of size $3\times 3$ the input matrix is partitioned in to sub-matrices of size $3\times 3$ and as the input matrix, the scheme $1$ is used to sustain computation to communication ratio. In implementation of dgeqrfggr, we update trailing matrix using dgemm while in implementation of dgeqr2ggr, the trailing matrix is updated using DET2 instructions. Attained speed-up over sequential realization in different Tile array sizes and matrix sizes is shown in figure \ref{fig:parallel_perf}. Speed-up in dgeqr2ggr, dgeqr2, dgeqrf, dgeqrfht, and dgeqr2ht over sequential realizations of these routines. It can be observed in the figure \ref{fig:ht_graph_5} that the speed-up attained in dgeqr2ggr is commensurate with the Tile array size in REDEFINE. For a Tile array size of $K\times K$, speed-up asymptotically approaches $K\times K$ as depicted in the figure \ref{fig:ht_graph_5}. Performance of dgeqr2ggr, dgeqrfggr, dgeqr2, dgeqrf, dgeqrfht, and dgeqr2ht in-terms of theoretical peak performance of Tile array size is shown in figure \ref{fig:ht_graph_6}. It can be observed in the figure \ref{fig:ht_graph_6} that dgeqrf2ggr can attain up to 78\% of the theoretical peak performance in REDEFINE for different Tile array sizes. 

%% file: conclusion.tex
Generalization of Givens Rotation was presented that resulted in lower multiplication count compared to classical Givens Rotation operation. Generalized Givens Rotation was implemented on multicore and General Purpose Graphics Processing Units where the performance was limited due to inability of these platforms in exploiting available parallelism in the routine. It was proposed to move away from traditional software packages based approach to architectural customizations for Dense Linear Algebra computations. Several macro operations were identified in Generalized Givens Rotation and realized on a Reconfigurable Data-path that is tightly coupled to pipeline of a Processing Element. Generalized Givens Rotation outperformed Modified Householder Transform presented in the literature by 10\% in Processing Element where Modified Householder Transform is implemented with similar approach of algorithm-architecture co-design. For parallel realization, the Processing Element was attached to REDEFINE Coarse-grained Reconfigurable Architecture as a Custom Function Unit and scalibility of the solution was shown where speed-up in parallel realization asymptotically approaches Tile array size in REDEFINE. 